\documentclass[%
 reprint,
 amsmath,amssymb,
 aps,
floatfix
]{revtex4-2}

\usepackage{graphicx}
\usepackage{dcolumn}
\usepackage{bm}
\usepackage[mathlines]{lineno}

\graphicspath{{./images/}}
\begin{document}


\title{Dramatic Failure of the Callaway Description of Heat Flow in Boron Arsenide and Boron Antimonide Driven by Phonon Scattering Selection Rules}

\author{Nikhil Malviya}
\author{Navaneetha K. Ravichandran}
\email{navaneeth@iisc.ac.in}
\affiliation{%
 Department of Mechanical Engineering, Indian Institute of Science, Bangalore 560012, India
 }%

\date{\today}

\begin{abstract}
Callaway's simplified heat flow model is often used to confirm experimental realizations of unconventional, hydrodynamic and Poiseuille phonon transport in ultrahigh thermal conductivity ($\kappa$) materials, due to its simplicity and low computational cost. Here, we show that the Callaway model works exceptionally well for most ultrahigh-$\kappa$ materials like diamond and boron nitride, but fails dramatically for boron arsenide (BAs) and boron antimonide (BSb). This failure is driven by the inability of the Callaway model to effectively describe the severely restricted phonon scattering in BAs and BSb, where many scattering selection rules are activated simultaneously. Our work highlights the powerful predictive capability of the Callaway model, and gives insights into the nature of phonon scattering in ultrahigh-$\kappa$ materials and the suitability of the Callaway's description of heat flow through them.
\end{abstract}

\maketitle
\clearpage

\section{Introduction}
Thermal transport in ultrahigh thermal conductivity ($\kappa$) materials has become a topic of considerable interest~\cite{tian_unusual_2018, kang_experimental_2018, li_high_2018, chen_ultrahigh_2020, li_anomalous_2022}, owing to their importance in the development of passive, energy efficient heat spreaders for low thermal noise and high power microelectronics~\cite{kang_integration_2021, graham_integrating_2021}. Additionally, recent experiments have shown that these materials are excellent platforms to realize unconventional non-diffusive hydrodynamic heat transport, apart from the conventional diffusive heat flow governed by the Fourier's law~\cite{lee_hydrodynamic_2015, machida_phonon_2020, huberman_observation_2019, ding_observation_2022}, which could open up possibilities for thermal cloaking and shielding of sensitive electronics.

In these materials, heat is carried by phonons, which are the quantized eigenmodes of the harmonic part of the crystal Hamiltonian. The remaining perturbative anharmonic terms in the crystal Hamiltonian drive two different types of scattering processes among phonons - those conserving the total quasimomentum of the participating phonons, called Normal (N) processes, and those dissipating a part of it to the crystal lattice, called Umklapp (U) processes. In the ultrahigh-$\kappa$ materials, the N-processes are much stronger than the U-processes~\cite{pitaevskii_physical_2012, krumhansl_thermal_1965, guyer_solution_1966, hardy_phonon_1970}, resulting in very low resistance to heat flow and enabling strong collective hydrodynamic phonon transport in them.

Phonon transport in these materials is governed by the Peierls Boltzmann equation (PBE)~\cite{peierls_zur_1929} given by:
\begin{equation}
    \label{eq:PBE}
    \frac{\partial n_{\lambda}}{\partial t} 
    + \textbf{v}_{\lambda} \cdot \mathbf{\nabla} n_{\lambda} 
    = \mathcal{C} \left(n_{\lambda} \right)
\end{equation}
where $n_{\lambda}$ is the non-equilibrium distribution function of a phonon mode $\lambda \equiv \left(\textbf{q}, j \right)$, with wave-vector $\textbf{q}$ and polarization $j$ at a temperature $T$ and time $t$, $\textbf{v}_{\lambda}$ is the phonon group velocity and $\mathcal{C} \left(n_{\lambda} \right)$ is the collision integral, describing the rate at which $n_{\lambda}$ changes in space and time due to phonon scattering processes. The PBE (eq.~\ref{eq:PBE}), even in its linearized version, strongly couples the non-equilibrium distribution functions of any two phonon modes~\cite{ward_ab_2009, lindsay_first-principles_2013, feng_four-phonon_2017, cepellotti_thermal_2016, feng_four-phonon_2018}, particularly for ultrahigh-$\kappa$ materials, thus making it challenging to gain qualitative insights on phonon transport from its solution. Hence, several previous works have used a simplified version of $\mathcal{C} \left(n_{\lambda} \right)$, originally proposed by Joseph Callaway~\cite{callaway_model_1959}, to obtain a microscopic perspective of phonon transport in a number of ultrahigh-$\kappa$ materials such as diamond~\cite{ma_examining_2014}, graphite~\cite{ding_phonon_2018}, graphene sheets~\cite{cepellotti_phonon_2015, li_role_2018}, graphane, boron nitride (BN), fluorographene, molybdenum disulphide~\cite{cepellotti_phonon_2015} and black phosphorus~\cite{ding_umklapp_2018}, under both diffusive and hydrodynamic conditions. However, it is unclear if the simplified Callaway description of heat flow is universally applicable for \emph{all} materials.

Here we show that, while the Callaway description of heat flow works exceedingly well for most ultrahigh-$\kappa$ materials, it fails dramatically for two ultrahigh-$\kappa$ materials: boron arsenide (BAs) and boron antimonide (BSb) from 150 K till 1000 K, with a room temperature (RT) error of 26\% ($\approx$ 350 Wm$^{-1}$K$^{-1}$) and 23\% ($\approx$ 150 Wm$^{-1}$K$^{-1}$) on their $\kappa$'s respectively. We show that the unusually strong and simultaneous activation of multiple phonon scattering selection rules results in severely restricted phonon scattering in BAs and BSb, which cannot be captured by the collision integral in the Callaway model, by construction. Our results highlight the powerful predictive capability of the Callaway model for most ultrahigh-$\kappa$ materials, elucidate the unconventional nature of heat flow in BAs and BSb, and provide computationally inexpensive guidelines to quickly identify the suitability of the Callaway model for newly discovered materials in the future.

To check the validity of the Callaway approximation for different materials, we construct the solution of the linearized form of the PBE (eq.~\ref{eq:PBE}) by expanding $n_{\lambda}$ around the Bose-Einstein equilibrium distribution function $n^0_{\lambda}$ as $n_{\lambda} \approx n_{\lambda}^{0} + n_{\lambda}^{0} \left( n_{\lambda}^{0} + 1\right) \tilde{n}_{\lambda}^{1}$. Further, by assuming a steady-state, one-dimensional temperature gradient and the spatial gradients of the deviation $n_{\lambda} - n^0_{\lambda}$ to be negligible, the linearized PBE (LPBE) becomes~\cite{peierls_zur_1929, broido_lattice_2005, fugallo_ab_2013}:
\begin{equation}
    \label{eq:LPBE}
    v_{\lambda, x} \frac{\mathrm{d} T}{\mathrm{d} x} 
    \frac{\partial n_{\lambda}^{0}}{\partial T} 
    \approx \mathcal{C}^{\text{lin.}} \left(n_{\lambda}\right) = \sum_{\lambda'} \mathcal{R}_{\lambda \lambda'} \tilde{n}_{\lambda}^{1}
\end{equation}
where $\mathcal{C}^{\text{lin.}} \left(n_{\lambda}\right)$ is the linearized collision integral and $\mathcal{R}_{\lambda\lambda'}$ is the linearized collision matrix. In this work, we include the terms corresponding to the three-phonon and four-phonon processes in $\mathcal{R}_{\lambda\lambda'}$. The expressions for $\mathcal{C} \left(n_{\lambda} \right)$, $\mathcal{C}^{\text{lin.}} \left(n_{\lambda}\right)$ and $\mathcal{R}_{\lambda \lambda'}$, and the details of the first-principles methodology to solve eq.~\ref{eq:LPBE} are provided in Appendix - I and in Ref.~\cite{ravichandran_phonon-phonon_2020} (results for naturally occurring materials including phonon-isotope scattering are included in the supplementary section S1). A material's $\kappa$ is then obtained by solving the LPBE (eq.~\ref{eq:LPBE}) for $\tilde{n}_{\lambda}^{1}$ and using it in the heat flux expression as:
\begin{equation}
J_x = \frac{1}{\Omega}\sum_\lambda \hbar\omega_\lambda v_{\lambda, x} n_{\lambda}^{0} \left( n_{\lambda}^{0} + 1 \right) \tilde{n}^1_\lambda = -\kappa\frac{\mathrm{d}T}{\mathrm{d}x}   \label{eq:linear_response}
\end{equation}
where $\omega_\lambda$ is the phonon frequency and $\Omega$ is the crystal volume. It is instructive to note that $\mathcal{C}^{\text{lin.}}\left(n_\lambda\right)$ can be written as a sum of a diagonal and an off-diagonal part:
\begin{equation}
    \label{eq:PBE_RTA_nonRTA}
     \mathcal{C}^{\text{lin.}} \left(n_{\lambda}\right) = \sum_{\lambda'} \mathcal{R}_{\lambda \lambda'} \tilde{n}_{\lambda'}^{1}
    = \mathcal{R}_{\lambda}^{\left(0 \right)} \tilde{n}_{\lambda}^{1} + 
    \sum_{\lambda'} \mathcal{R}_{\lambda \lambda'}^{\left(1 \right)} \tilde{n}_{\lambda'}^{1} 
\end{equation}
A commonly used approximation to the LPBE, the relaxation time approximation (RTA), is obtained by ignoring the off-diagonal terms of $\mathcal{C}^{\text{lin.}} \left(n_{\lambda}\right)$ [i.e., $\sum_{\lambda'} \mathcal{R}_{\lambda \lambda'}^{\left(1 \right)} \tilde{n}_{\lambda'}^{1}$] in eq.~\ref{eq:PBE_RTA_nonRTA}. The solution of the LPBE under the RTA provides simple insights into the phonon decay processes and their effect on $\kappa$, since there is no coupling among $n_\lambda$'s of different phonon modes in the RTA. However, $\kappa$ derived from the RTA ($\kappa_{\mathrm{RTA}}$) significantly under-predicts the complete solution of the LPBE for ultrahigh-$\kappa$ materials, since the collision integral under the RTA, given by:
\begin{equation}
    \label{eq:RTA_approx_explained}
    \mathcal{C}^{\text{RTA}} \left(n_{\lambda}\right)
    = \mathcal{R}_{\lambda}^{\left(0 \right)} \tilde{n}_{\lambda}^{1} = -\frac{n_{\lambda} - n_{\lambda}^{0}}{\tau_{\lambda}^{\text{U}}}
    -\frac{n_{\lambda} - n_{\lambda}^{0}}{\tau_{\lambda}^{\text{N}}}
\end{equation}
assumes that both the N- and the U-processes drive $n_\lambda$ towards $n^0_\lambda$ with rates $1/\tau_{\lambda}^{\text{N}}$ and $1/\tau_{\lambda}^{\text{U}}$ respectively. However, in the absence of U-processes, the N-processes drive $n_\lambda$ towards $n^*_\lambda = \frac{1}{\exp \left( \frac{\hbar \omega_{\lambda} }{k_{B}T} + \mathbf{\Lambda}\cdot\textbf{q} \right) - 1} \approx \left[n_{\lambda}^{0} - n_{\lambda}^{0} \left(n_{\lambda}^{0} + 1 \right) \textbf{q} \cdot \mathbf{\Lambda}\right]$, a drifting equilibrium distribution function with the constant $\mathbf{\Lambda}$ related to the phonon mobility $\mathbf{\Theta}$ as $\mathbf{\Lambda} = \mathbf{\Theta}\frac{\text{d}T}{\text{d}x}$~\cite{cepellotti_phonon_2015}. Hence, when the scattering rates of the N-processes ($1/\tau^{\text{N}}$) are much stronger than those of the U-processes ($1/\tau^{\text{U}}$), as in ultrahigh-$\kappa$ materials, the RTA (eq.~\ref{eq:RTA_approx_explained}) fails to capture their high $\kappa$.

To overcome this problem, Callaway \cite{callaway_model_1959} proposed an improved approximation to $\mathcal{C}^{\text{lin.}} \left(n_{\lambda}\right)$ for ultrahigh-$\kappa$ materials, where the dissipative U-processes relax $n_\lambda$ towards $n_{\lambda}^{0}$, while the non-dissipative N-processes relax $n_\lambda$ towards $n_{\lambda}^{*}$. The LPBE (eq.~\ref{eq:LPBE}) under the Callaway approximation becomes:
\begin{equation}
\label{eq:PBE_callaway}
    v_{\lambda, x} \frac{\mathrm{d} T}{\mathrm{d} x} 
    \frac{\partial n_{\lambda}^{0}}{\partial T}  = 
    - \frac{n_{\lambda} - n_{\lambda}^{0}}{\tau_{\lambda}^{\text{U}}} 
    - \frac{n_{\lambda} - n_{\lambda}^{*}}{\tau_{\lambda}^{\text{N}}} = \mathcal{C}^{\text{Call.}}\left(n_\lambda\right)
\end{equation}
thus maintaining the uncoupled nature of the collision integral with respect to $n_\lambda$. The phonon mobility $\mathbf{\Theta}$, needed to calculate $n_{\lambda}^{*}$ in eq.~\ref{eq:PBE_callaway}, is determined using an additional closure condition on the exact conservation of phonon quasimomentum $\hbar\mathbf{q}$ in the presence of N-processes only, i.e.,
\begin{equation}
\sum_\lambda \hbar\mathbf{q}\mathcal{C}^{\text{lin., N}}\left(n_\lambda\right) = \sum_{\lambda\lambda'}\hbar\mathbf{q}\mathcal{R}^{\text{N}}_{\lambda\lambda'}\tilde{n}_{\lambda'}^{1} = 0    \label{eq:Callaway_closure}
\end{equation}
Here, $\mathcal{C}^{\text{lin., N}}\left(n_\lambda\right)$ and $\mathcal{R}^{\text{N}}_{\lambda\lambda'}$ are the linearized collision integral and matrix respectively, in the presence of N-processes only. Identifying $\tilde{n}^{*1}_{\lambda'} = \left[n^*_\lambda - n^0_\lambda\right]/\left[n^0_\lambda\left(n^0_\lambda + 1\right)\right]$ as belonging to the null space of $\mathcal{R}^{\text{N}}_{\lambda\lambda'}$~\cite{krumhansl_thermal_1965, guyer_solution_1966} and neglecting the off-diagonal terms in $\mathcal{R}^{\text{N}}_{\lambda\lambda'}$, we obtain the closure condition originally used by Callaway~\cite{callaway_model_1959} as:
\begin{equation}
\sum_{\lambda\lambda'}\hbar\mathbf{q}\mathcal{R}^{\text{N}}_{\lambda\lambda'}\left[\tilde{n}_{\lambda'}^{1} - \tilde{n}^{*1}_{\lambda'}\right] \approx -\sum_\lambda \hbar\mathbf{q}\frac{n_{\lambda} - n^*_{\lambda}}{\tau_{\lambda}^{\text{N}}} = 0 \label{eq:Callaway_closure_RTA}
\end{equation}
Assuming linearized forms of $n_\lambda$ and $n_{\lambda}^{*}$, eqs.~\ref{eq:linear_response},~\ref{eq:PBE_callaway} and~\ref{eq:Callaway_closure_RTA} can be solved to obtain $\kappa$ from the Callaway model ($\kappa_{\text{Call.}}$) as:
\begin{align}
\label{eq:kappa_callaway}
    \kappa_{\text{Call.}} 
    = &
    \frac{1}{\Omega} \sum_{\lambda} \hbar \omega_{\lambda} \tau_{\lambda}^{\text{T}} v_{\lambda, x}^{2} \frac{\partial n_{\lambda}^{0}}{\partial T} \nonumber \\ &
    + \frac{1}{\Omega} \sum_{\lambda} \frac{\tau_{\lambda}^{\text{T}}}{\tau_{\lambda}^{\text{N}}} \hbar \omega_{\lambda} v_{\lambda, x} n_{\lambda}^{0} \left(n_{\lambda}^{0} + 1 \right) \textbf{q} \cdot \mathbf{\Theta}
\end{align}
Here, $\tau_{\lambda}^{T}$ is the total phonon relaxation time, obtained using the Matthiessen's rule from the relaxation times of N- and U-processes as $1/\tau_{\lambda}^{\text{T}} = 1/\tau_{\lambda}^{\text{N}} + 1/\tau_{\lambda}^{\text{U}}$, and the phonon mobility $\mathbf{\Theta}$ is given by:
\begin{equation*}
    \mathbf{\Theta} 
    = \left[ 
    \sum_{\lambda} \textbf{q} \otimes \textbf{q} 
    \frac{\tau_{\lambda}^{\text{T}}}{\tau_{\lambda}^{\text{N}} \tau_{\lambda}^{\text{U}}} 
    n_{\lambda}^{0} \left(n_{\lambda}^{0} + 1 \right) 
    \right]^{-1} 
    \left(
    \sum_{\lambda} \textbf{q} v_{\lambda, x} 
    \frac{\tau_{\lambda}^{\text{T}}}{\tau_{\lambda}^{\text{N}}} 
    \frac{\partial n_{\lambda}^{0}}{\partial T} 
    \right)
\end{equation*}
where $\otimes$ represents a dyadic product (for a detailed derivation, see Appendix - I).

\section{Results and discussion}
\begin{figure}[!ht]
\includegraphics[width = 0.95\linewidth]{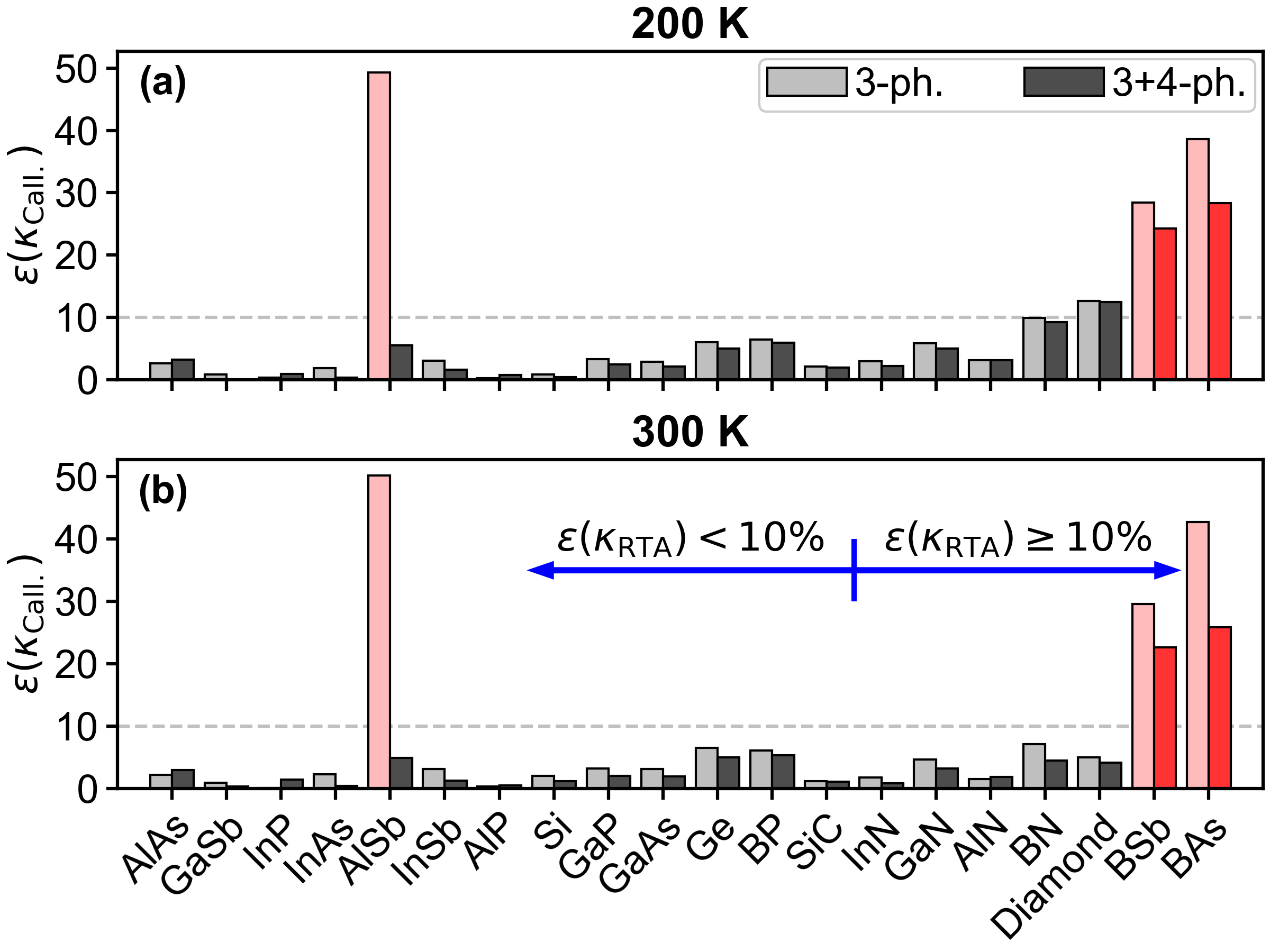}
\caption{\label{fig:kappa_relative_error} Percentage error in $\kappa_{\text{Call.}}$, $\epsilon \left(\kappa_{\text{Call.}} \right) = 100 \times \left\vert1 - \frac{\kappa_{\text{Call.}}}{\kappa_{\text{LPBE}}} \right\vert$ \%, for twenty materials at 200 K and 300 K, with (dark color bars) and without (light color bars) the inclusion of four-phonon scattering. The materials listed to the right of silicon carbide (SiC) have a percentage error of more than 10\% in $\kappa_{\text{RTA}}$ including four-phonon scattering.}
\end{figure}

Figures~\ref{fig:kappa_relative_error} (a) and (b) show the error in $\kappa_{\text{Call.}}$ [$\epsilon \left(\kappa_{\text{Call.}} \right)$] relative to that obtained from the LPBE ($\kappa_{\text{LPBE}}$) for twenty different cubic semiconductors with varying masses of the constituent atoms and strengths of their inter-atomic bonds. Three important observations can be made from this figure. First, the Callaway model predicts, with reasonable accuracy [$\epsilon\left(\kappa_{\text{Call.}}\right) < 10\%$], the $\kappa$ at 200 K and 300 K of those materials where the RTA description of phonon transport including three-phonon and four-phonon scattering is sufficient. For these materials, the distinction between N- and U-processes is not important as their N- and U-scattering rates are comparable. Therefore the predictions of the Callaway model are in agreement with those of the RTA, thus confirming our expectation. Second, it also accurately predicts the $\kappa$ of certain ultrahigh-$\kappa$ materials like diamond and BN at 200 K and 300 K, where the RTA is known to fail dramatically~\cite{ward_ab_2009, chen_ultrahigh_2020}. The percentage errors [$\epsilon\left(\kappa_{\text{Call.}}\right)$] are only about 12\% and 4\% for diamond, and 9\% and 5\% for BN at 200 K and 300 K respectively, including four-phonon scattering. Third, in stark contrast to the above two cases, the Callaway model fails dramatically for two other ultrahigh-$\kappa$ materials - BAs and BSb, at 200 K and 300 K, with or without the inclusion of four-phonon scattering in our calculations. For BAs, $\epsilon\left(\kappa_{\text{Call.}}\right)$ is about 26\% (28\%) at 300 K (200 K), while for BSb, it is about 23\% (24\%) at 300 K (200 K) including four-phonon scattering. Since the Callaway model works for aluminum antimonide (AlSb) when four-phonon scattering is included, we do not discuss this case further (see Appendix - II for details).

\begin{figure*}[!ht]
\includegraphics[width = 0.99\linewidth]{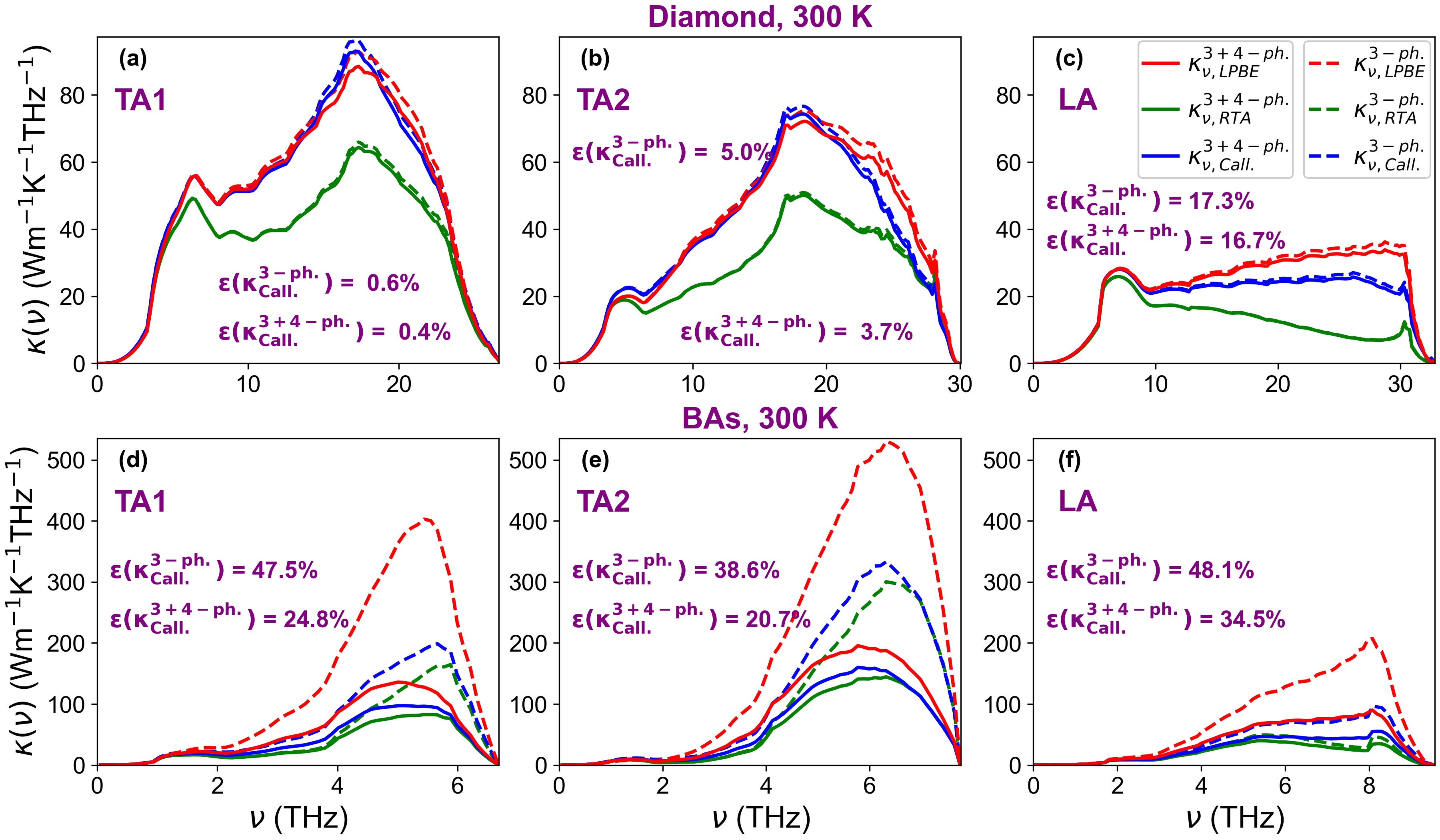}
\caption{\label{fig:iso_kappa_Diamond_BAs} Modewise (TA1, TA2 and LA phonon modes) spectral contributions of $\kappa$ for diamond and BAs at 300K, with (solid)  and without (dashed) the inclusion of four-phonon scattering.}
\end{figure*}

Interestingly, for diamond, the agreement between the Callaway and the LPBE solution is much better for the first [0.4\% (7\%)] and second [4\% (10\%)] transverse acoustic branches - TA1 and TA2 respectively, than for the longitudinal acoustic - LA branch [17\% (27\%)] at 300 K (200 K) as shown in Fig.~\ref{fig:iso_kappa_Diamond_BAs} for 300 K. Similar excellent agreement between the Callaway and the LPBE solutions is observed for the TA1 and TA2 branches in BN as well. However, since the overall contribution to $\kappa$ is much larger from the TA1 and TA2 branches than from the LA branch, the Callaway solution closely approximates that of the LPBE in diamond and BN. For both diamond and BN, the RTA significantly under-predicts the $\kappa$ for TA1, TA2 and LA branches relative to the LPBE solution at 200 K and 300 K.

To understand why the Callaway model works so well for diamond and BN in general, and for TA1 and TA2 phonons in these materials in particular, but fails dramatically for BAs and BSb, we recast $\mathcal{C}^{\text{Call.}}\left(n_\lambda\right)$ into a form similar to $\mathcal{C}^{\text{lin.}}\left(n_\lambda\right)$ in eq.~\ref{eq:PBE_RTA_nonRTA}. To get to this form, we first solve eq.~\ref{eq:Callaway_closure_RTA} for $\mathbf{\Lambda}$ using the linearized forms of $n_{\lambda}$ and $n_{\lambda}^{*}$ as:
\begin{gather}
    \sum_{\lambda} \hbar \textbf{q} \left( \frac{n_{\lambda}^{0} \left(n_{\lambda}^{0}+1 \right) \tilde{n}_{\lambda}^{1}}{\tau_{\lambda}^{\text{N}}} + \frac{n_{\lambda}^{0} \left(n_{\lambda}^{0}+1 \right) \textbf{q} \cdot \mathbf{\Lambda}}{\tau_{\lambda}^{\text{N}}} \right) = 0 \nonumber
\end{gather}
Thus,
\begin{gather}
\mathbf{\Lambda} = -\sum_{\lambda} \left( \frac{\tilde{\textbf{q}} n_{\lambda}^{0} \left(n_{\lambda}^{0} + 1 \right) \tilde{n}_{\lambda}^{1}}{\tau_{\lambda}^{\text{N}}} \right) \\
    \text{where\ }
    \tilde{\textbf{q}} = \left[ \sum_{\lambda'} \textbf{q}' \otimes \textbf{q}'\frac{ n_{\lambda'}^{0}(n_{\lambda'}^{0}+1)}{\tau_{\lambda'}^{\text{N}}} \right]^{-1}\textbf{q} \label{eq:callaway_lambda}
\end{gather}
Next, we substitute the resulting $\mathbf{\Lambda}$ into eq.~\ref{eq:PBE_callaway}, with linearized forms of $n_{\lambda}$ and $n_{\lambda}^{*}$, to get the collision matrix that corresponds to $\mathcal{C}^{\text{Call.}} \left( n_{\lambda} \right)$ as (see Ref.~\cite{nettleton_foundations_1963} and Appendix - I for details):
\begin{align}
    \mathcal{C}^{\text{Call.}}\left(n_\lambda\right) 
    \approx &
    - \frac{n_{\lambda} - n_{\lambda}^{0}}{\tau_{\lambda}^{\text{U}}}
    - \frac{n_{\lambda} - n_{\lambda}^{0} + n_{\lambda}^{0} \left(n_{\lambda}^{0} + 1 \right) \textbf{q} \cdot \mathbf{\Lambda}}{\tau_{\lambda}^{\text{N}}} \nonumber \\
    = & 
    - \frac{n_{\lambda}^{0} \left(n^{0}_{\lambda}+1 \right) \tilde{n}_{\lambda}^{1}}{\tau_{\lambda}^{\text{T}}}
    - \frac{n_{\lambda}^{0} \left(n_{\lambda}^{0} + 1 \right) \textbf{q} \cdot \mathbf{\Lambda}}{\tau_{\lambda}^{\text{N}}} \nonumber \\
    = & 
    \mathcal{R}_{\lambda}^{\left(0 \right)}\tilde{n}_{\lambda}^{1} + \sum_{\lambda'} \mathcal{S}^{\left(1 \right)}_{\lambda \lambda'}\tilde{n}_{\lambda'}^{1}   \label{eq:Callaway_colmatdef}\\
    \text{with  }\mathcal{S}^{\left(1 \right)}_{\lambda \lambda'} &= \textbf{q} \cdot \tilde{\textbf{q}}' \frac{n_{\lambda}^{0} \left(n_{\lambda}^{0} + 1 \right)}{\tau_{\lambda}^{\text{N}}} \frac{n_{\lambda'}^{0} \left(n_{\lambda'}^{0} + 1 \right)}{\tau_{\lambda'}^{\text{N}}}\label{eq:Callaay_S_offdiagonal}
\end{align}
Here, $\mathcal{S}^{\left(1\right)}_{\lambda\lambda'}$ are the off-diagonal elements of the Callaway collision matrix $\mathcal{S}_{\lambda\lambda'}$. The Callaway collision integral is constructed in such a way that the diagonal terms of $\mathcal{R}_{\lambda\lambda'}$ and $\mathcal{S}_{\lambda\lambda'}$ are both equal to $\mathcal{R}^{\left(0\right)}_\lambda$. Thus, the Callaway model will poorly approximate the LPBE solution when (1) the off-diagonal contribution to $\mathcal{C}^{\text{lin.}}\left(n_\lambda\right)$, i.e., $\sum_{\lambda'} \mathcal{R}_{\lambda \lambda'}^{\left(1 \right)} \tilde{n}_{\lambda'}^{1}$ is comparable to or larger than the diagonal term, i.e., $\mathcal{R}_{\lambda}^{\left(0 \right)} \tilde{n}_{\lambda}^{1}$, and (2) the off-diagonal matrix elements $\mathcal{R}^{\left(1\right)}_{\lambda\lambda'}$ and $\mathcal{S}^{\left(1\right)}_{\lambda\lambda'}$ are very different.

\begin{figure}[!ht]
\includegraphics[width = 0.95\linewidth]{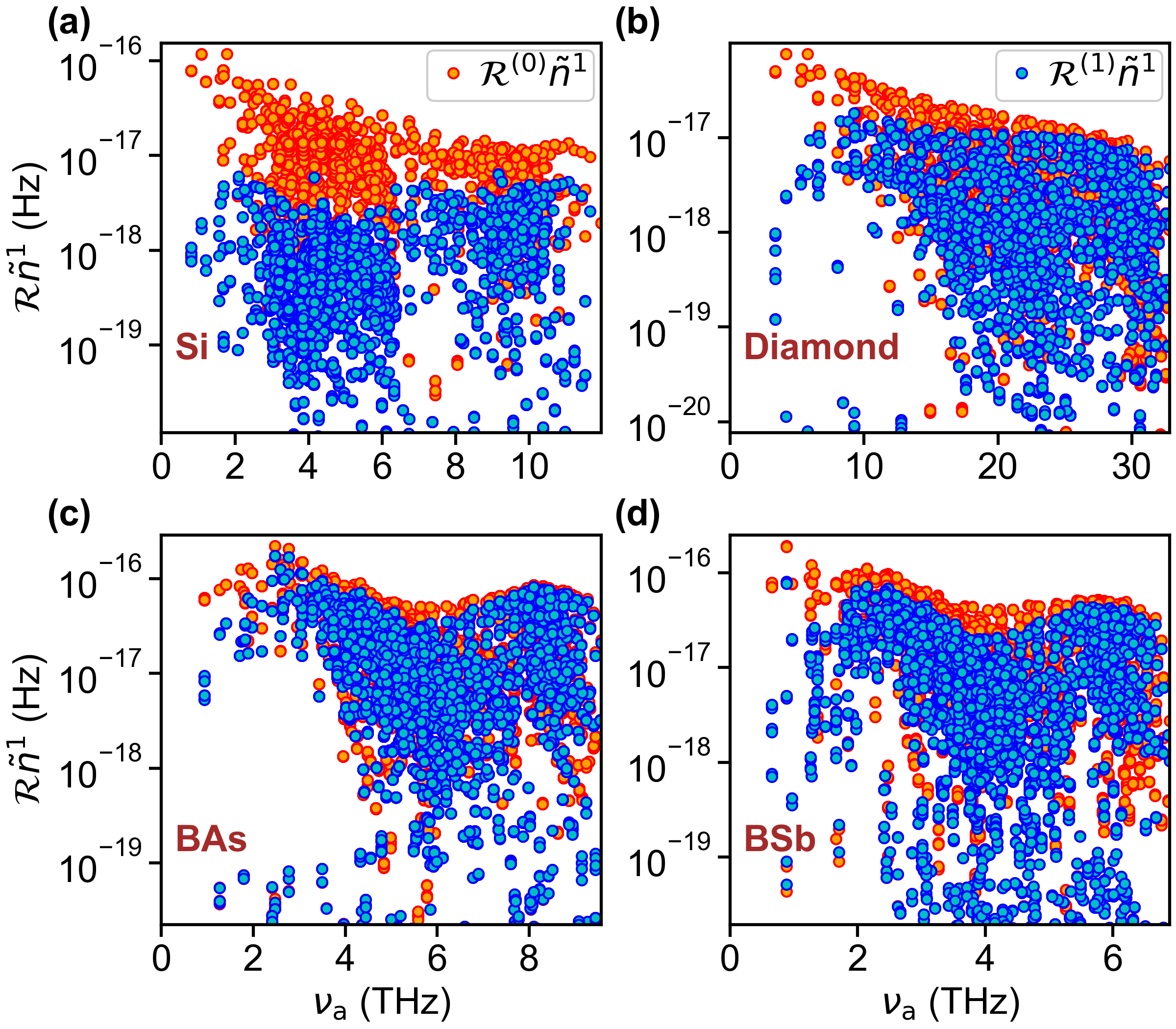}
\caption{\label{fig:rta_validation_plot} Comparison of the diagonal and off-diagonal parts of the collision integral in LPBE for Si (a), diamond (b), BAs (c) and BSb (d) at 300 K as a function of phonon frequency $\nu_a$.}
\end{figure}
Figure~\ref{fig:rta_validation_plot} shows the comparison between the diagonal and the off-diagonal contributions to $\mathcal{C}^{\text{lin.}}\left(n_\lambda\right)$ for the acoustic phonons in different materials at 300 K. For materials like Si, where $\kappa_{\text{RTA}} \approx \kappa_{\text{LPBE}}$, the diagonal terms are much larger than the off-diagonal terms [Fig.~\ref{fig:rta_validation_plot} (a)]. As shown in Fig. S4 in the supplementary section S2, the diagonal terms dominate over the off-diagonal terms even for $\mathcal{C}^{\text{Call.}}\left(n_\lambda\right)$ in Si. Hence, for Si, $\kappa_{\text{RTA}} \approx \kappa_{\text{LPBE}} \approx \kappa_{\text{Call.}}$. On the other hand, for materials like diamond, BAs and BSb [Fig.~\ref{fig:rta_validation_plot} (b), (c) and (d) respectively], the diagonal and the off-diagonal terms of $\mathcal{C}^{\text{lin.}}\left(n_\lambda\right)$ are comparable, thus satisfying the first failure condition of the Callaway model.

\begin{figure}[!ht]
\includegraphics[width = 0.95\linewidth]{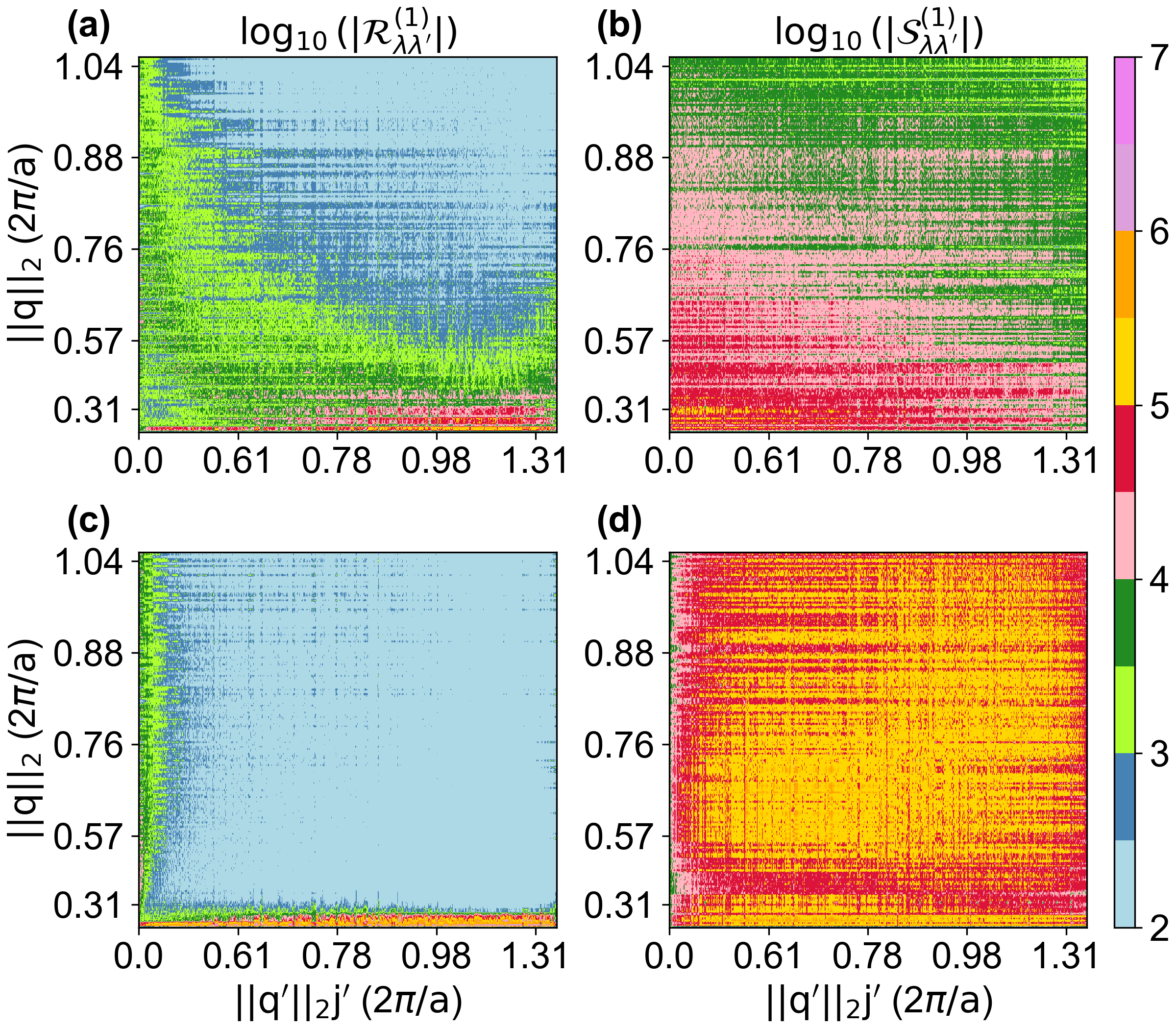}
\caption{\label{fig:collision_matrix} The off-diagonal collision matrices from the LPBE ($\mathcal{R}^{\left(1\right)}_{\lambda\lambda'}$) and the Callaway model ($\mathcal{S}^{\left(1\right)}_{\lambda\lambda'}$) for the TA1 phonon branch in diamond [(a) and (b) respectively] and the LA phonon branch in BAs [(c) and (d) respectively]. The ordinate lists the $\mathbf{q}$-points (in units of $2\pi/a$, where $a$ is the lattice constant) of the TA1 (LA) phonons only, while the abscissa has six points for each unique $\mathbf{q}'$, corresponding to the six polarizations in diamond (BAs).} 
\end{figure}

To check the second failure condition for the Callaway model in diamond, we focus on the TA1 polarization, for which the Callaway solution closely approaches the LPBE solution at 300 K, while the RTA solution deviates strongly [see Fig.~\ref{fig:iso_kappa_Diamond_BAs} (a)]. In Figs.~\ref{fig:collision_matrix} (a) and (b), we compare $\mathcal{R}^{\left(1\right)}_{\lambda\lambda'}$ and $\mathcal{S}^{\left(1\right)}_{\lambda\lambda'}$ for the TA1 phonons in diamond at 300 K. Although $\mathcal{S}^{\left(1\right)}_{\lambda\lambda'}$ is simplistic, it qualitatively captures the important features of $\mathcal{R}^{\left(1\right)}_{\lambda\lambda'}$ accurately. Specifically, the features of large $\mathcal{R}^{\left(1\right)}_{\lambda\lambda'}$ for small $\|\mathbf{q}\|_2$ or small $\|\mathbf{q}'\|_2$ where the N-processes dominate, and the vanishingly small $\mathcal{R}^{\left(1\right)}_{\lambda\lambda'}$ for large $\|\mathbf{q}\|_2$ and large $\|\mathbf{q}'\|_2$ where the U-processes dominate, as shown in Fig.~S9 in the supplementary section S3, are also reflected in $\mathcal{S}^{\left(1\right)}_{\lambda\lambda'}$. Thus, the LPBE picture of strong N-scattering and relatively weak U-scattering in diamond at 300 K, which is the origin of the large enhancement of $\kappa$ from the RTA to the complete LPBE solution, is qualitatively captured by the Callaway collision matrix. The small quantitative differences in $\mathcal{R}^{\left(1\right)}_{\lambda\lambda'}$ and $\mathcal{S}^{\left(1\right)}_{\lambda\lambda'}$ do not affect the results significantly, since the off-diagonal terms of $\mathcal{C}^{\text{lin.}}\left(n_\lambda\right)$ and $\mathcal{C}^{\text{Call.}}\left(n_\lambda\right)$ are still smaller than the diagonal terms for the vast majority of the TA1 phonons. Similar features are also observed in the collision matrix of BN, as shown in Fig.~S10 in the supplementary section S3.

\begin{figure}[!ht]
\includegraphics[width = 0.90\linewidth]{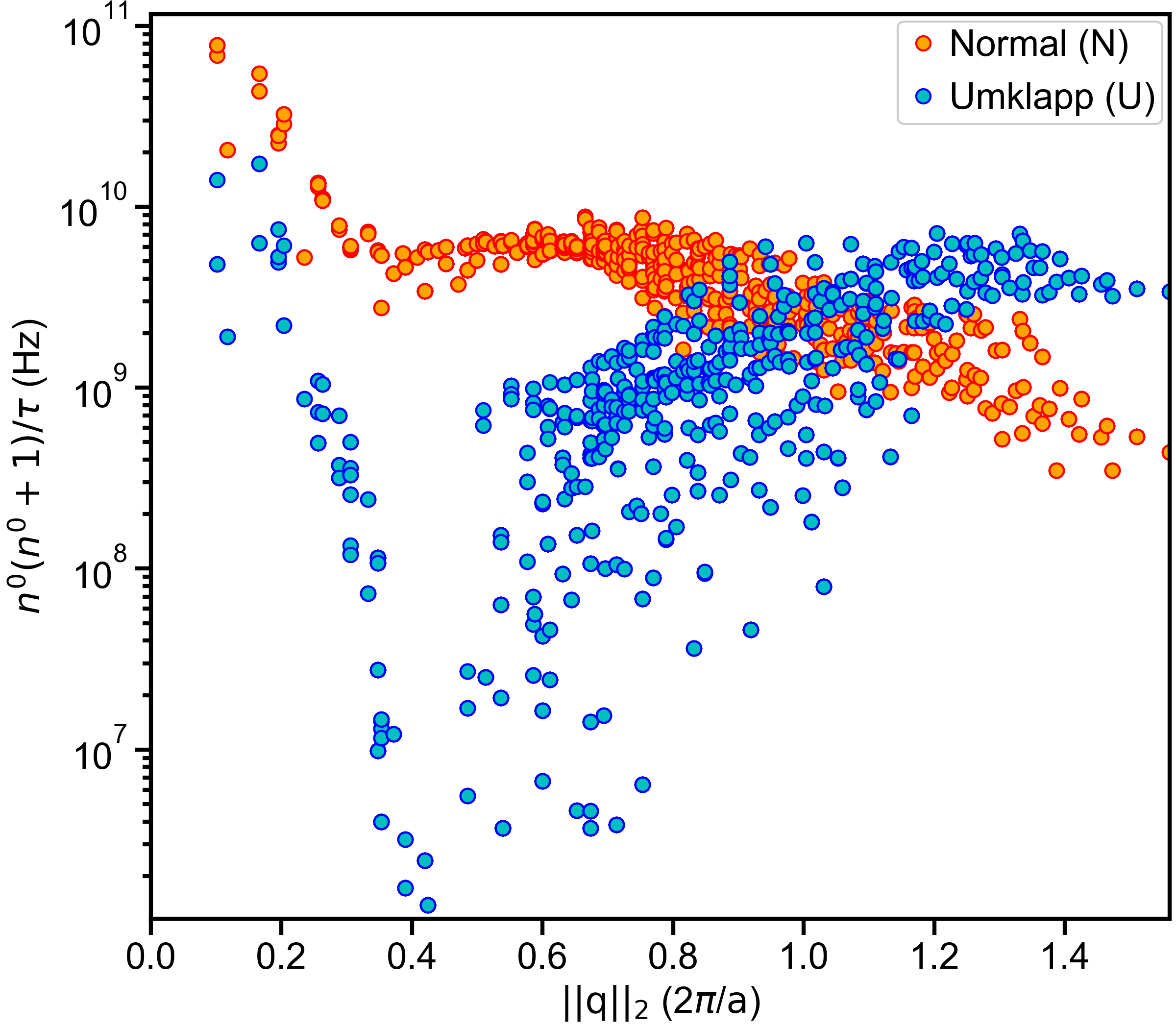}
\caption{\label{fig:bas_scattering_rate} Comparison of $\mathcal{R}^{\left(\text{N/U}, 0\right)}_\lambda = n^0_\lambda\left(n^0_\lambda + 1\right)/\tau^{\left(\text{N/U}\right)}_\lambda$ for the N- and the U-processes of LA phonons in BAs at 300 K.}
\end{figure}

On the other hand, $\mathcal{S}^{\left(1\right)}_{\lambda\lambda'}$ fails to capture the features of $\mathcal{R}^{\left(1\right)}_{\lambda\lambda'}$ in BAs at 300 K. In BAs, the acoustic phonons are bunched together, which severely restricts the lowest-order three-phonon interactions among them~\cite{lax_spontaneous_1981, lindsay_first-principles_2013, ravichandran_phonon-phonon_2020}. Additionally, the large mass difference between the boron and arsenic atoms results in a large frequency gap between the acoustic and the optic phonons, that almost completely forbids any three-phonon interactions among them. The simultaneously strong activation of these three-phonon scattering selection rules~\cite{ravichandran_phonon-phonon_2020} results in vanishingly small values of $\mathcal{R}^{\left(1\right)}_{\lambda\lambda'}$ for a large number of $\left(\lambda, \lambda'\right)$ pairs except when $\|\mathbf{q}\|_2$ or $\|\mathbf{q}'\|_2$ is small, as shown in Fig.~\ref{fig:collision_matrix} (c) for the LA phonons in BAs, for which the error $\epsilon \left( \kappa_{\text{Call.}}\right)$ is the largest as shown in Fig.~\ref{fig:iso_kappa_Diamond_BAs} (see Fig.~S11 in the supplementary section S3 for the $\mathcal{R}^{\left(1\right)}_{\lambda\lambda'}$ contours of the TA1 and TA2 phonons in BAs). As shown in Fig.~S9 in the supplementary section S3, the region of large $\|\mathbf{q}\|_2$ and $\|\mathbf{q}'\|_2$ in these contours is dominated by the U-processes due to the momentum conservation restriction, and so, the vanishingly small $\mathcal{R}^{\left(1\right)}_{\lambda\lambda'}$ for BAs in this region explains the overall weak U-scattering of acoustic phonons with large frequencies, as observed earlier~\cite{ravichandran_phonon-phonon_2020} and also represented in Fig.~\ref{fig:bas_scattering_rate}.

\begin{figure}[!ht]
    \centering
    \includegraphics[width=0.99\linewidth]{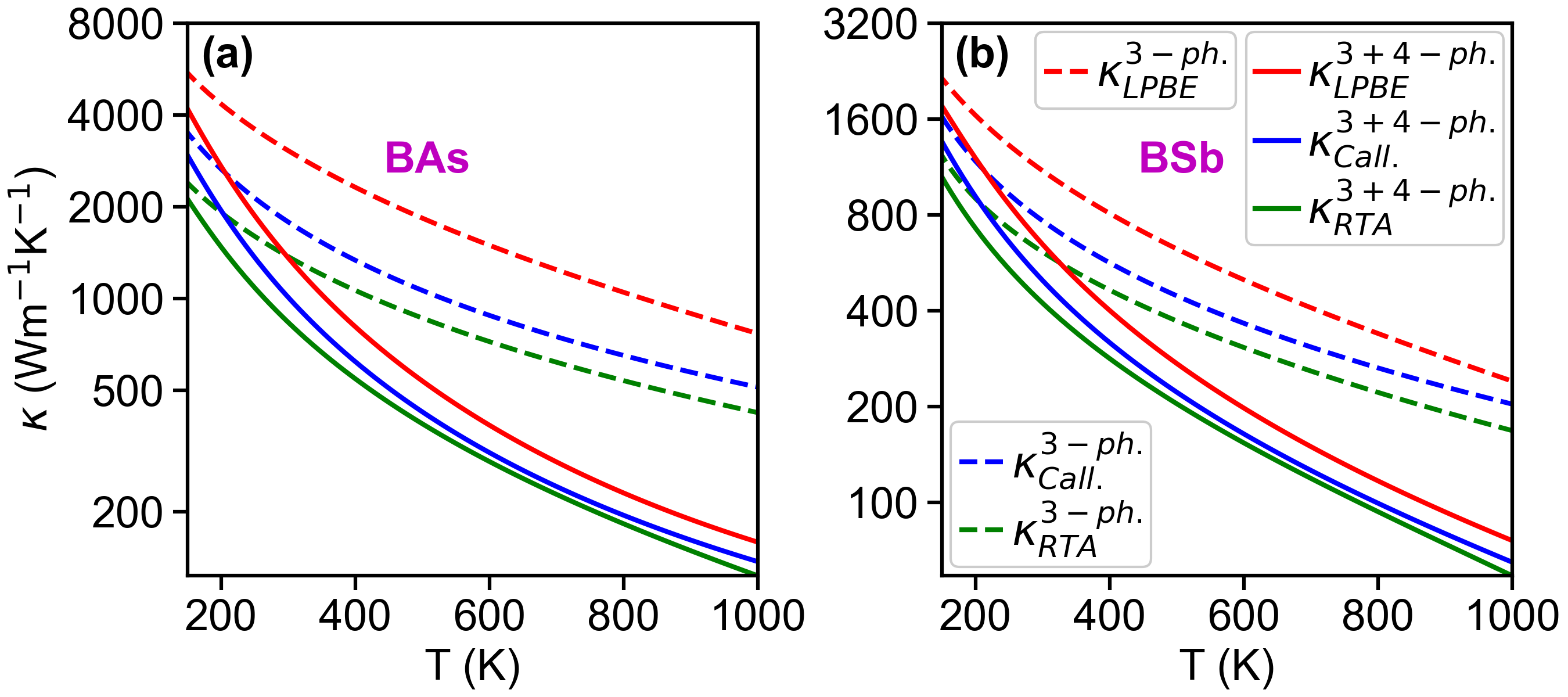}
    \caption{Thermal conductivity ($\kappa$) of BAs (a) and BSb (b) at different temperatures (T), calculated by solving the complete LPBE (red), the LPBE under the RTA (green), and the LPBE using the Callaway model (blue), with (solid line) and without (dashed line) the inclusion of four-phonon scattering.}
    \label{fig:kappa_temp}
\end{figure}

In stark contrast, Fig.~\ref{fig:collision_matrix} (d) shows the exact opposite trend in $\mathcal{S}^{\left(1\right)}_{\lambda\lambda'}$, with more than an order of magnitude larger $\mathcal{S}^{\left(1\right)}_{\lambda\lambda'}$ when $\|\mathbf{q}\|_2$ and $\|\mathbf{q}'\|_2$ are large, compared to the case when $\|\mathbf{q}\|_2$ or $\|\mathbf{q}'\|_2$ is small. This behavior originates from the absence of any information on U-processes in the expression for $\mathcal{S}^{\left(1\right)}_{\lambda\lambda'}$ (eq.~\ref{eq:Callaay_S_offdiagonal}), since it depends on the relaxation times of the N-processes only. As shown in Fig.~\ref{fig:bas_scattering_rate}, the factor $n^0_\lambda\left(n^0_\lambda + 1\right)/\tau_{\lambda}^{\text{N}}$ for the N-processes, which appears in $\mathcal{S}^{\left(1\right)}_{\lambda\lambda'}$, is weakly dependent on $\|\mathbf{q}\|_2$. Therefore, for a fixed $\|\mathbf{q}\|_2$, $\mathcal{S}^{\left(1\right)}_{\lambda\lambda'}$ does not decay fast enough as $\|\mathbf{q}'\|_2$ increases in Fig.~\ref{fig:collision_matrix} (d), to capture the sharply decreasing $\mathcal{R}^{\left(1\right)}_{\lambda\lambda'}$ with increasing $\|\mathbf{q}'\|_2$ in Fig.~\ref{fig:collision_matrix} (c). Similar trends in $\mathcal{R}^{\left(1\right)}_{\lambda\lambda'}$ and $\mathcal{S}^{\left(1\right)}_{\lambda\lambda'}$ are also found for BSb, as shown in the supplementary sections S2 and S3. Thus, due to the complete qualitative and quantitative misrepresentation of $\mathcal{R}^{\left(1\right)}_{\lambda\lambda'}$ by $\mathcal{S}^{\left(1\right)}_{\lambda\lambda'}$, the Callaway model fails dramatically for BAs and BSb at 300 K. In fact, the Callaway model performs well for diamond and BN over a broad temperature range as shown in the supplementary section S4, but it is unable to predict the LPBE solution for $\kappa$ in BAs and BSb anywhere between 150 K to 1000 K, as shown in Figs.~\ref{fig:kappa_temp}~(a) and~(b) respectively (similar plots for the other eighteen materials, with and without the inclusion of phonon-isotope scattering, are included in the supplementary section S4).

\begin{figure}[ht!]
    \centering
    \includegraphics[width=0.95\linewidth]{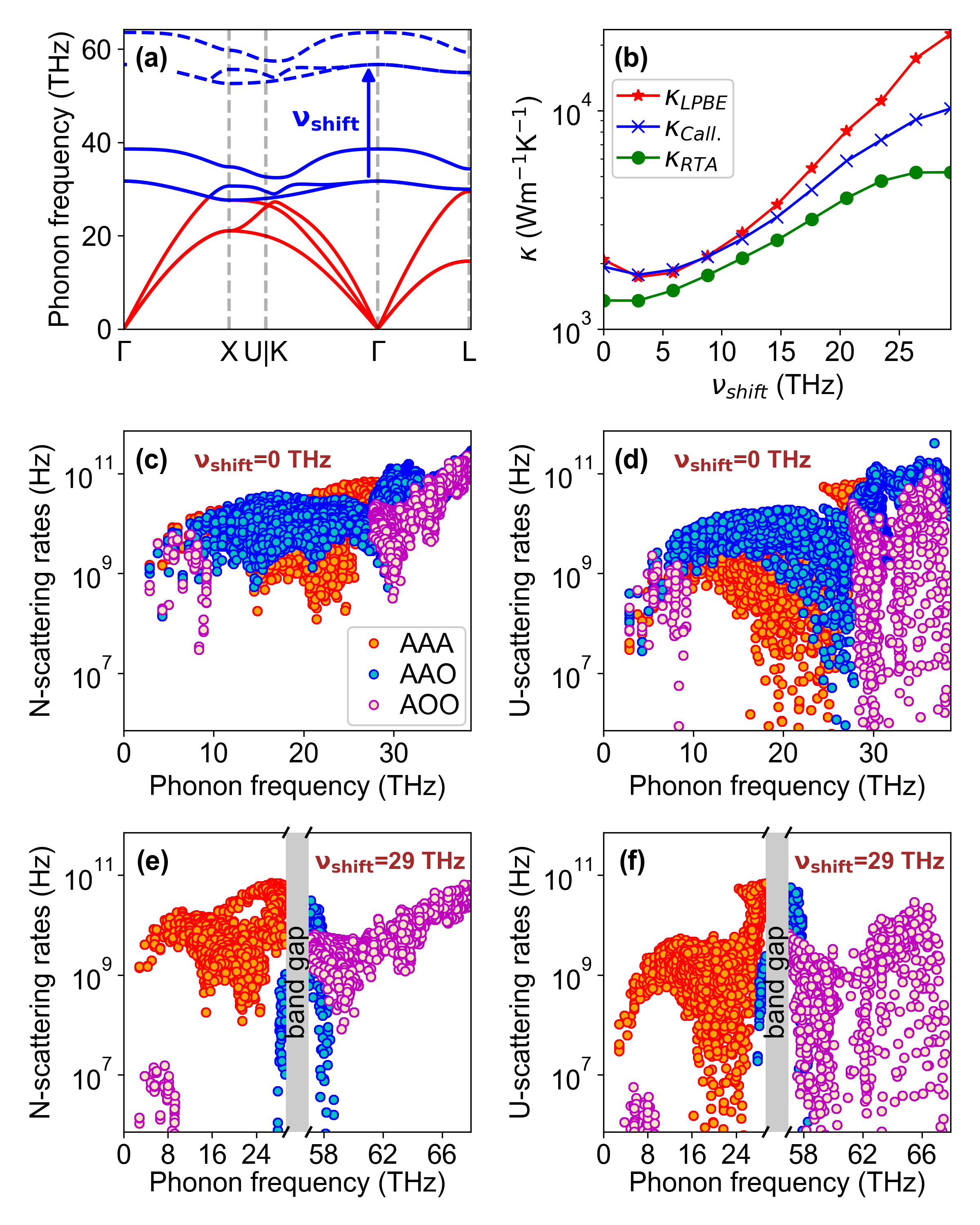}
    \caption{A computational experiment showing the failure of the Callaway model by artificially activating multiple three-phonon scattering selection rules in boron nitride (BN) at 300K. (a) Phonon dispersions for BN showing the acoustic (red) and the optic (blue) phonon branches. The optic branches are systematically shifted to create and widen the A-O band gap. (b) $\kappa$ calculated using the LPBE (red), the Callaway model (blue) and the RTA (green), as a function of the shift in the optic branches. Figures (c) and (d) show the three-phonon N- and U-scattering rates respectively, for BN without any shift in the optic branches. Figures (e) and (f) show the three-phonon N- and U-scattering rates respectively for BN with a shift in the frequencies of the optic branches equal to the acoustic bandwidth. The same harmonic and anharmonic force constants are used for all of these calculations.}
    \label{fig:systematic_bn}
\end{figure}

It is important to note here that the failure of the Callaway model is due to the \emph{simultaneous} activation of multiple three-phonon scattering selection rules. To further highlight the importance of this requirement, we conducted a computational experiment on BN at 300 K. BN exhibits small optic bandwidth and acoustic bunching, but the A-O (A: acoustic, O: optic) band gap vanishes since the basis atoms, boron and nitrogen, have nearly the same masses. In our computational experiment, we artificially shifted the optic branches to higher frequencies systematically, and computed the $\kappa$, as depicted in Figs.~\ref{fig:systematic_bn}~(a) and~(b). To keep the calculations simple, we included scattering among three phonons only for this computational experiment, since the effects of four-phonon scattering on the $\kappa$ of BN is weak at 300 K and there are no selection rules for these higher-order scattering processes, as discussed in Refs.~\cite{ravichandran_phonon-phonon_2020, chen_ultrahigh_2020}. Initially, when there is no shift, significant AAO scattering is observed,  that compensates for the low AAA U-scattering, as shown in Figs.~\ref{fig:systematic_bn}~(c) and~(d). In this case, the phonon-phonon scattering processes are not highly restricted, and so, the Callaway model provides a reasonable estimate of $\kappa$ compared to the complete solution of the LPBE. As we increase the band gap from zero to fifteen THz, we do not observe any notable differences between the $\kappa$ values obtained using the LPBE and the Callaway model, since the low AAA scattering is always compensated by the large AAO scattering within this range of the optic phonon frequency shift. However, when the A-O gap exceeds fifteen THz, the low AAA U-scattering rates begin to manifest, and they are fully exposed as the A-O gap approaches the acoustic bandwidth, as shown in Figs.~\ref{fig:systematic_bn}~(e) and~(f). In this range, the error in the prediction of $\kappa$ from the Callaway model relative to the complete solution of the LPBE progressively worsens as the band gap increases, as shown in Fig.~\ref{fig:systematic_bn}~(b). Thus, the \emph{simultaneous} activation of multiple phonon scattering selection rules that results in a highly restricted phonon scattering phase space is a necessary condition for the failure of the Callaway model in predicting the $\kappa$ of a material.

\section{Modifications to the Callaway model}
Over the past few years, two variations to the original Callaway model have been proposed in the literature. In the first variant~\cite{allen_improved_2013}, which we refer to as the Allen's modified Callaway model, a different quasimomentum conservation condition has been used to obtain $\mathbf{\Lambda}$, by enforcing the condition that the total quasimomentum of phonons remains invariant when only N-scattering among phonons occurs in a crystal. This condition is represented by:
\begin{gather}
    \label{eq:allen_condition_for_momentum_conservation}
    \sum_{\lambda} \hbar \textbf{q} \left( n_{\lambda} - n^{*}_{\lambda} \right) = 0 
\end{gather}
Using the quasimomentum conservation condition as in eq.~\ref{eq:allen_condition_for_momentum_conservation}, the phonon mobility for the calculation of $\kappa$ and the off-diagonal elements of the collision matrix in the Allen's modified Callaway model are obtained as: 
\begin{align}
    \mathbf{\Theta}
    = & \left[ 
    \sum_{\lambda} \textbf{q} \otimes \textbf{q} n_{\lambda}^{0} \left(n_{\lambda}^{0} + 1 \right) \frac{\tau_{\lambda}^{\text{T}}}{\tau_{\lambda}^{\text{U}}} \right]^{-1} \nonumber \\ &
    \left( \sum_{\lambda} \textbf{q} \left( v_{\lambda, x} \tau_{\lambda}^{T} \frac{\partial n_{\lambda}^{0}}{\partial T} \right) \frac{\mathrm{d} T}{\mathrm{d} x} \right)
\end{align}
\begin{gather}
    \mathcal{S}^{ \left(1 \right)}_{\lambda \lambda'}
    = \textbf{q} \cdot \tilde{\textbf{q}}' \frac{n_{\lambda}^{0} \left(n_{\lambda}^{0} + 1 \right)}{\tau_{\lambda}^{N}}  n_{\lambda'}^{0} \left(n_{\lambda'}^{0} + 1 \right) \label{Eq:Allen_Collision}\\
    \text{where, } 
    \tilde{\textbf{q}} = \left[ 
    \sum_{\lambda} \textbf{q} \otimes \textbf{q} n_{\lambda}^{0} \left(n_{\lambda}^{0} + 1 \right) \right]^{-1} \textbf{q} \nonumber
\end{gather}

\begin{figure}[ht!]
    \centering
    \includegraphics[width=0.95\linewidth]{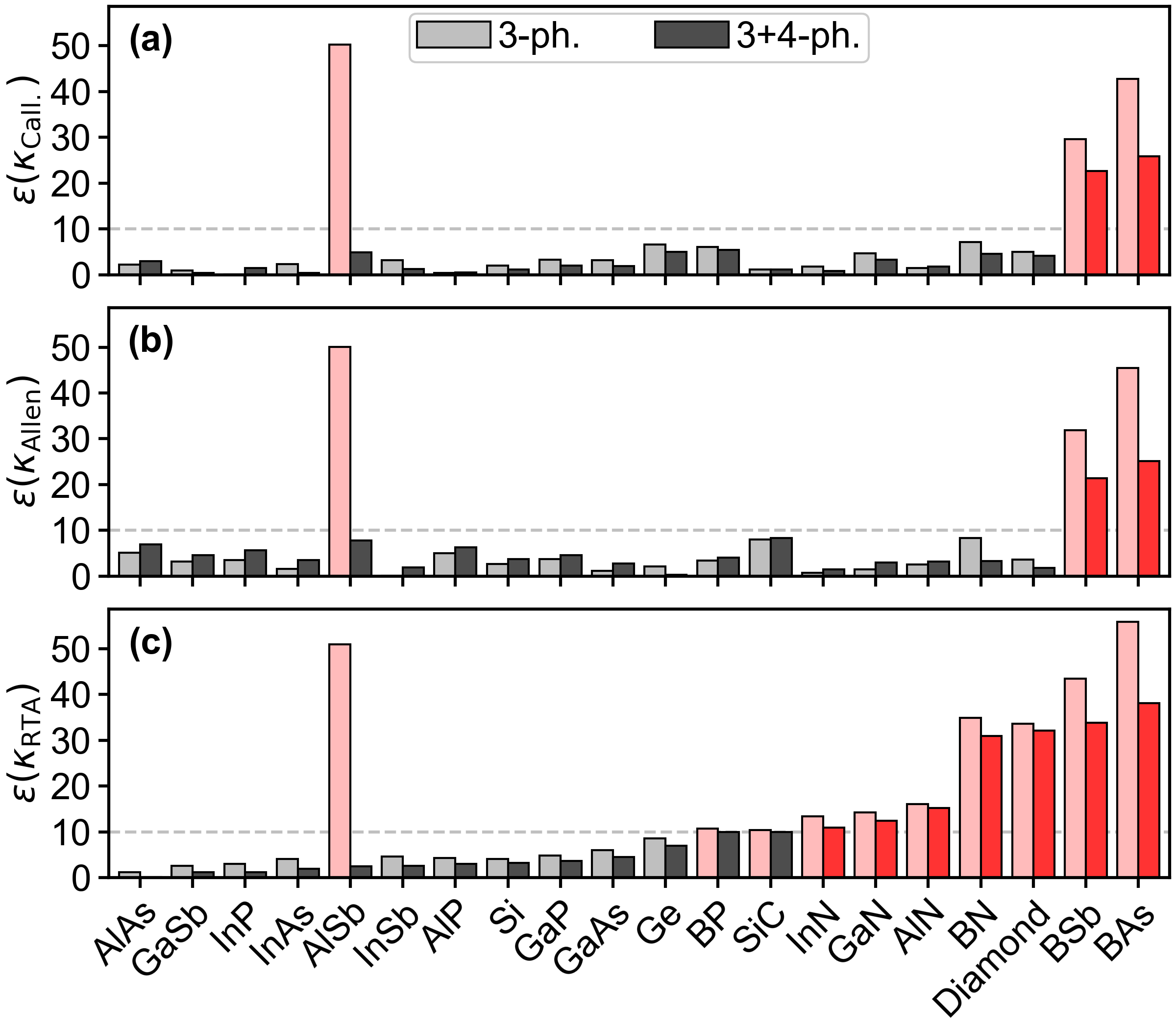}
    \caption{Percentage error in $\kappa$ - $\epsilon \left( \kappa \right) = 100 \times \left\vert 1 - \frac{\kappa}{\kappa_{\text{LPBE}}} \right\vert$ calculated using (a) the Callaway model, (b) the Allen's modified Callaway model, and (c) the RTA, for twenty materials at 300 K, with (dark color bars) and without (light color bars) including four-phonon scattering. The materials listed here are sorted according to increasing $\epsilon \left( \kappa_{\text{RTA}} \right)$ with the inclusion of four-phonon scattering.}
    \label{fig:other_callaway_kappa_error_bar_plot_sorted_4ph_rta}
\end{figure}

\begin{figure}[!ht]
\includegraphics[width = 0.95\linewidth]{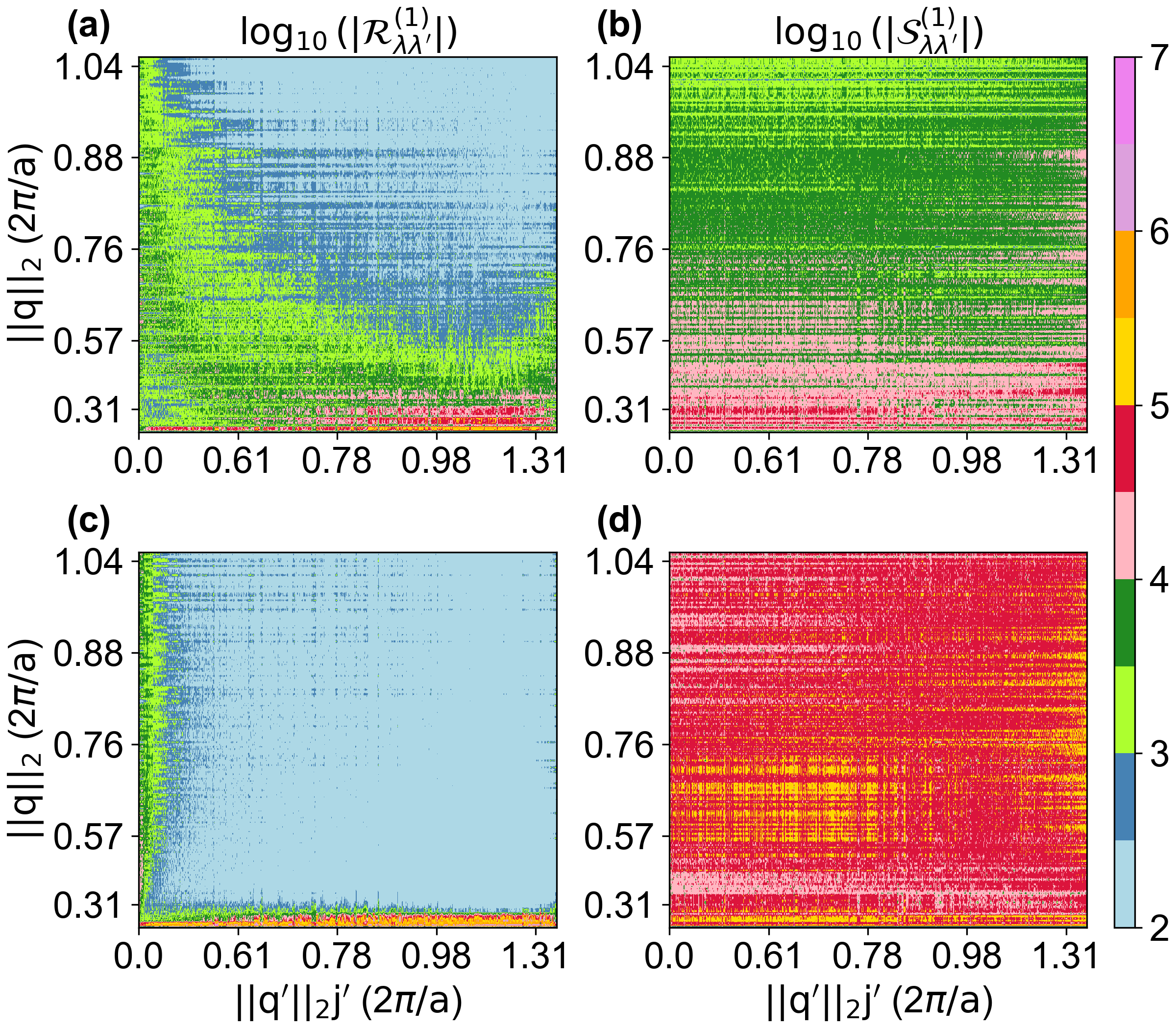}
\caption{\label{fig:allen_contour_subplot} The off-diagonal collision matrices from the LPBE ($\mathcal{R}^{\left(1\right)}_{\lambda\lambda'}$) and the Allen's modified Callaway model ($\mathcal{S}^{\left(1\right)}_{\lambda\lambda'}$) for the TA1 phonon branch in diamond [(a) and (b) respectively] and the LA phonon branch in BAs [(c) and (d) respectively]. The ordinate lists the $\mathbf{q}$-points (in units of $2\pi/a$, where $a$ is the lattice constant) of the TA1 (LA) phonons only, while the abscissa has six points for each unique $\mathbf{q}'$, corresponding to the six polarizations in diamond (BAs).} 
\end{figure}

Figure~\ref{fig:other_callaway_kappa_error_bar_plot_sorted_4ph_rta} shows the error in $\kappa$ ($\epsilon \left( \kappa \right)$) calculated by the RTA, the Callaway and the Allen's modified Callaway models relative to that obtained from the solutions of the LPBE ($\kappa_{\text{LPBE}}$) for the twenty cubic materials studied here. We see from this figure that, for diamond and BN, the Callaway model, both in its original form and with Allen's modification to it, performs exceedingly well in predicting the complete solutions of the LPBE. However, both models severely fail for BAs and BSb. Once again, the failure of the Allen's modified Callaway model for BAs and BSb is due to its inability to replicate the highly restricted three-phonon scattering phase-space in these two materials, as shown in the contour plots for the LA phonons in BAs [Fig.~\ref{fig:allen_contour_subplot} (c) and (d)]. It is to be noted from Fig.~\ref{fig:allen_contour_subplot} (b) that, even though the contours of the collision matrix from Allen's modified Callaway model show the correct features of large $\mathcal{S}^{(1)}_{\lambda\lambda'}$ values at small $\left(\|\mathbf{q}\|_2, \|\mathbf{q}'\|_2\right)$ and small $\mathcal{S}^{(1)}_{\lambda\lambda'}$ values at large $\left(\|\mathbf{q}\|_2, \|\mathbf{q}'\|_2\right)$ for the TA1 phonons in diamond, the qualitative features at $\left[\text{small }\|\mathbf{q}\|_2, \text{large }\|\mathbf{q}'\|_2\right]$ and $\left[\text{large }\|\mathbf{q}\|_2, \text{small }\|\mathbf{q}'\|_2\right]$ are slightly different compared to the respective regions in the collision matrix of the LPBE [Fig.~\ref{fig:allen_contour_subplot} (a)], due to the inherent asymmetry of the collision matrix of Allen's modified Callaway model (eq.~\ref{Eq:Allen_Collision}). Nevertheless, these slight differences between the off-diagonal elements of the collision matrix from the LPBE and the Allen's modified Callaway model do not influence the results significantly, since the off-diagonal terms of $\mathcal{C}^{\text{lin.}}\left(n_\lambda\right)$ and $\mathcal{C}^{\text{Call.}}\left(n_\lambda\right)$ are still smaller than the diagonal terms for the vast majority of the TA1 phonons in diamond, similar to the original Callaway model as mentioned earlier.

Another modification to the Callaway model has been proposed in Ref.~\cite{ding_umklapp_2018}, where the authors have modified the method of classifying the phonon-phonon scattering processes as N- and U-processes. In this modification, any scattering event that conserves phonon quasimomentum in the direction of the temperature gradient (i.e. $\sum_{i=1}^3 q_{xi} = 0$ for a three-phonon scattering process with the temperature gradient along the $x$-direction) is classified as an N-process, while those that do not fulfill this condition are classified as U-processes. This definition can be used along with the closure condition of the original Callaway model (eq.~\ref{eq:Callaway_closure_RTA}) or the Allen's modified Callaway model (eq.~\ref{eq:allen_condition_for_momentum_conservation}) to obtain the $\kappa$ values. With the original Callaway model (eq.~\ref{eq:Callaway_closure_RTA}) for a one-dimensional temperature gradient (say along the $x$-direction) and with the modified definition of the N- and the U-processes, the quasimomentum conservation condition becomes:
\begin{align}
    \label{eq:gang_chen_callaway_condition_for_momentum_conservation}
    - \sum_{\lambda'} \hbar q_{x} \left( \frac{ n_{\lambda} - n_{\lambda}^{*} }{\tau_{\lambda}^{N}} \right) 
    = 0 
\end{align}
Following a similar procedure to get the phonon mobility $\mathbf{\Theta}$ as in the previous cases, with eq.~\ref{eq:gang_chen_callaway_condition_for_momentum_conservation} as the quasimomentum conservation condition, we get:
\begin{align}
    \label{eq:get_gang_chen_callaway_theta}
    \sum_{\lambda} \frac{q_{x}}{\tau_{\lambda}^{N}} \tau_{\lambda}^{T} v_{\lambda, x}  \frac{\partial n_{\lambda}^{0}}{\partial T} 
    = 
    \sum_{\lambda} \frac{q_{x}}{\tau_{\lambda}^{N}} \frac{\tau_{\lambda}^{T}}{\tau_{\lambda}^{U}} n_{\lambda}^{0} \left( n_{\lambda}^{0} + 1 \right) \textbf{q} \cdot \mathbf{\Theta} 
\end{align}
Equation~\ref{eq:get_gang_chen_callaway_theta} is a single equation with three unknowns, which cannot be solved in general. However, by assuming that the component of $\mathbf{\Theta}$ is non-zero only along the temperature gradient, we can obtain an expression for $\Theta_x$ as:
\begin{align}
    \label{eq:get_gang_chen_theta_x}
    \Theta_{x} = \frac{\sum_{\lambda} \frac{q_{x}}{\tau_{\lambda}^{N}} 
    \tau_{\lambda}^{T} v_{\lambda, x} \frac{\partial n_{\lambda}^{0}}{\partial T}}{\sum_{\lambda} \frac{q_{x}}{\tau_{\lambda}^{N}} \frac{\tau_{\lambda}^{T}}{\tau_{\lambda}^{U}} n_{\lambda}^{0} \left( n_{\lambda}^{0} + 1 \right) q_x} 
\end{align}
Similarly, $\Theta_{x}$ can be obtained with the Allen's modified condition for quasimomentum conservation (eq.~\ref{eq:allen_condition_for_momentum_conservation}) in a separate calculation.

\begin{figure}[ht!]
    \centering
    \includegraphics[width=0.95\linewidth]{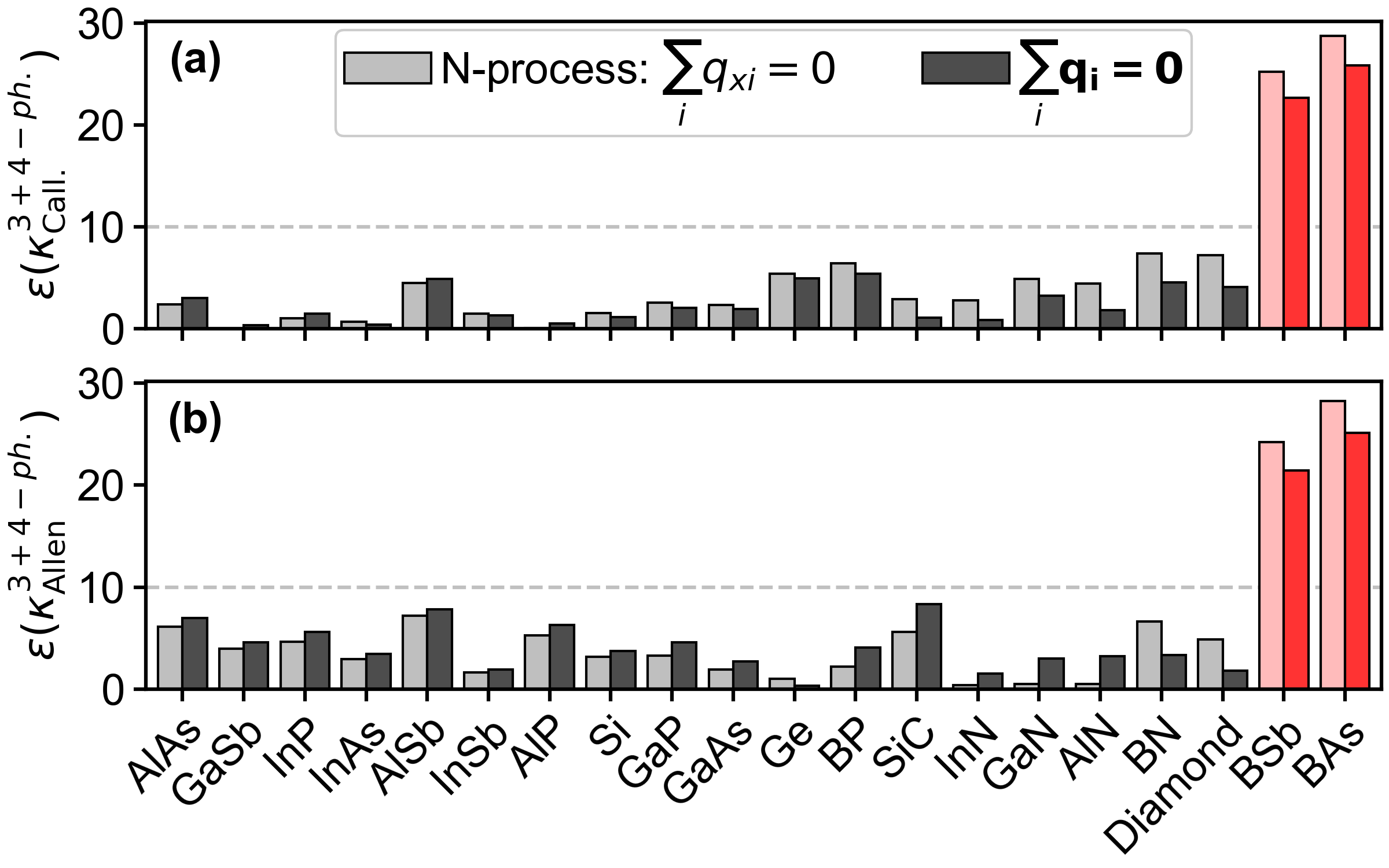}
    \caption{Percentage error in $\kappa$ - $\epsilon \left( \kappa \right) = 100 \times \left\vert 1 - \frac{\kappa}{\kappa_{\text{LPBE}}} \right\vert$ calculated using (a) the Callaway model, and (b) the Allen's modified Callaway model, for twenty materials at 300 K, including four-phonon scattering. Here the N-processes are defined in two different ways - the conventional definition which conserves the total quasimomentum of interacting phonons, i.e., $\sum_{i}\textbf{q}_{i}=\mathbf{0}$ (dark color bars) and a modified definition as in Ref.~\cite{ding_umklapp_2018}, which conserves the total quasimomentum of interacting phonons along the direction of the temperature gradient ($x$-direction here), i.e., $\sum_{i} q_{xi}=0$ (light color bars). In the above conditions, the sum over $i$ runs from one to three for three-phonon processes and from one to four for four-phonon processes.}
    \label{fig:new_other_callaway_kappa_error_bar_plot_sorted_old_4ph_rta}
\end{figure}

Figure~\ref{fig:new_other_callaway_kappa_error_bar_plot_sorted_old_4ph_rta} shows the comparison of $\epsilon \left( \kappa \right)$ from the original Callaway and the Allen's modified Callaway models, considering the conventional and the newly defined N- and the U-processes as in Ref.~\cite{ding_umklapp_2018}, with the inclusion of four-phonon scattering. Upon adopting the modified classification procedure for the N- and the U-processes as in Ref.~\cite{ding_umklapp_2018}, a few of the conventionally defined U-processes will now be categorized as N-processes. Since this reclassification does not affect the total scattering rates, the RTA outcomes remain unaffected and so, $\epsilon \left( \kappa \right)$ from the RTA is not compared in this figure. However, for both the original Callaway and the Allen's modified Callaway models, we find that $\epsilon \left( \kappa \right)$ is lower while adopting the conventional definition of the N- and U-processes than while using the newly defined N- and U-processes as in Ref.~\cite{ding_umklapp_2018}, particularly for ultrahigh-$\kappa$ materials. Most importantly, even with the newly defined N- and U-processes, both the original Callaway and the Allen's modified Callaway models fail to predict the $\kappa$ values of BAs and BSb, due to the additional reduction in the number of U-processes caused by this reclassification.\\

\section{Conclusion and Outlook}
In summary, we have shown that the Callaway description of heat flow works exceptionally well for most ultrahigh-$\kappa$ materials except BAs and BSb. This peculiar failure for BAs and BSb is caused by the inability of the Callaway collision matrix to capture the strong and \emph{simultaneous} activation of multiple phonon scattering selection rules, that restrict the allowed phonon decay processes in these two materials. Our work elucidates the unconventional nature of heat flow in BAs and BSb, compared to other ultrahigh-$\kappa$ materials like diamond and BN. Owing to the simplicity and low computational cost of the Callaway model compared to the complete solution of the LPBE, we anticipate that our work will inspire development of new computational tools, such as the Monte Carlo schemes for phonon transport~\cite{ravichandran_coherent_2014, peraud_efficient_2011, peraud_alternative_2012} using the Callaway model to directly simulate heat flow through complex nanoscale devices made of some of the ultrahigh-$\kappa$ materials like diamond and BN, as well as new Callaway-like simplifications for coupled transport problems involving phonon-spin and phonon-electron interactions~\cite{li_colossal_2022, li_high-frequency_2023, protik_electron-phonon_2020, protik_coupled_2020}.

\section{Acknowledgement}
This work was supported by the Core Research Grant (CRG) no. CRG/2020/006166 and the Mathematical Research Impact Centric Support (MATRICS) grant no. MTR/2022/001043 from the Department of Science and Technology - Science and Engineering Research Board, India. NM gratefully acknowledges the Prime Minister's Research Fellowship (PMRF) grant no. PMRF-02-01036. NR thanks the Infosys Foundation for their support through a Young Investigator Award. 

\begin{widetext}
\section*{Appendix - I: Methods}
\subsection{First-principles calculation of the phonon collision matrix}
In this work, we have considered three-phonon and four-phonon scattering processes. The total phonon collision integral ($\mathcal{C}\left(n_\lambda\right)$) for these processes can be written as:
\begin{align}
    \mathcal{C}\left(n_\lambda\right) = \mathcal{C}^{\text{3-ph.}} \left( n_{\lambda} \right) + \mathcal{C}^{\text{4-ph.}} \left( n_{\lambda} \right)
\end{align}
The individual collision integrals for these phonon scattering processes are expanded below.
\subsubsection{Three-phonon collision matrix}
The three-phonon collision integral $\mathcal{C}^{\text{3-ph.}} \left( n_{\lambda} \right)$ is obtained from the Fermi's golden rule as
\begin{align}
    \label{eq:3-ph_collosion_integral}
    \mathcal{C}^{\text{3-ph.}} \left( n_{\lambda} \right) 
    = & - \frac{\pi \hbar}{4N_{0}} \sum_{\lambda' \lambda''} 
    \Bigg[
    \frac{ \vert \Phi_{\lambda \lambda' \left( -\lambda''\right)} \vert^{2} }
    { \omega_{\lambda} \omega_{\lambda'} \omega_{\lambda''} }
    \left( 
    n_{\lambda} n_{\lambda'} \left( n_{\lambda''} + 1 \right) 
    - \left( n_{\lambda} + 1 \right) \left( n_{\lambda'} + 1 \right) n_{\lambda''}
    \right) \delta\left( \omega_{\lambda} + \omega_{\lambda'} - \omega_{\lambda''} \right) \nonumber \\
    &
    + \frac{1}{2}
    \frac{ \vert \Phi_{\lambda \left( -\lambda'\right) \left( -\lambda''\right)} \vert^{2} }
    { \omega_{\lambda} \omega_{\lambda'} \omega_{\lambda''} }
    \left( 
    n_{\lambda} \left( n_{\lambda'} + 1 \right) \left( n_{\lambda''} + 1 \right) 
    - \left( n_{\lambda} + 1 \right) n_{\lambda'} n_{\lambda''}
    \right) \delta\left( \omega_{\lambda} - \omega_{\lambda'} - \omega_{\lambda''} \right)
    \Bigg]
\end{align}
 Here, $N_{0}$ is the number of unit cells in the crystal. The three-phonon matrix elements $\mathbf{\Phi}_{\lambda \lambda' \lambda''}$ are related to the third order inter-atomic force constants $\Phi_{\alpha \beta \gamma} \left( lk,l'k',l''k'' \right)$ as:
\begin{align}
    \label{eq:fourier_transformed_force_constant}
    \mathbf{\Phi}_{\lambda \lambda' \lambda''} 
    =
    \sum_{l'l''} \sum_{kk'k''} \sum_{\alpha \beta \gamma}  &
    \frac{\Phi_{\alpha \beta \gamma} \left( 0k,l'k',l''k'' \right)}{\sqrt{m_{k} m_{k'} m_{k''}}} w_{\alpha}\left( \textbf{q}j, k \right) w_{\beta}\left( \textbf{q}'j', k' \right) w_{\gamma}\left( \textbf{q}''j'', k'' \right) \nonumber \\
    &
    \exp{\left( i \textbf{q}' \cdot \textbf{R}\left( l'\right) \right)}
    \exp{\left( i \textbf{q}'' \cdot \textbf{R}\left( l''\right) \right)}
    \Delta\left( \textbf{q}+\textbf{q}'+\textbf{q}'' \right)
\end{align}
where $l, l', l''$ are the lattice site indices with position vectors - $\textbf{R}\left( l \right), \textbf{R}\left( l' \right)$ and $\textbf{R}\left( l'' \right)$ respectively, $k, k', k''$ are the indices for the types of basis atoms with masses $m_{k}$, $m_{k'}$ and $m_{k''}$ respectively, $\alpha, \beta, \gamma$ are the indices for the Cartesian coordinates, and $w$ is the phonon eigenvector.

Equation~\ref{eq:3-ph_collosion_integral}, upon linearization using $n_\lambda = n^0_\lambda + n^0_\lambda\left(n^0_\lambda + 1\right)\tilde{n}^1_\lambda$ becomes: 
\begin{align}
    \mathcal{C}^{\text{lin., 3-ph.}} \left( n_{\lambda} \right) 
    = & 
    - \frac{\pi \hbar}{4N_{0}}
    \sum_{\lambda' \lambda''} 
    \Bigg[
    \frac{ \vert \mathbf{\Phi}_{\lambda \lambda' \left(-\lambda''\right)} \vert^{2}}{\omega_{\lambda} \omega_{\lambda'} \omega_{\lambda''}} \delta\left( \omega_{\lambda} + \omega_{\lambda'} - \omega_{\lambda''}\right) n_{\lambda}^{0} n_{\lambda'}^{0} \left( n_{\lambda''}^{0} + 1 \right) \left( \tilde{n}_{\lambda}^{1} + \tilde{n}_{\lambda'}^{1} - \tilde{n}_{\lambda''}^{1} \right) \nonumber \\
    & + \frac{1}{2}
    \frac{ \vert \mathbf{\Phi}_{\lambda \left( -\lambda' \right) \left(-\lambda''\right)} \vert^{2}}{\omega_{\lambda} \omega_{\lambda'} \omega_{\lambda''}} \delta\left( \omega_{\lambda} - \omega_{\lambda'} - \omega_{\lambda''}\right) n_{\lambda}^{0} \left( n_{\lambda'}^{0} + 1 \right) \left( n_{\lambda''}^{0} + 1 \right) \left( \tilde{n}_{\lambda}^{1} - \tilde{n}_{\lambda'}^{1} - \tilde{n}_{\lambda''}^{1} \right) 
    \Bigg] \nonumber 
\end{align}

Using the short-hand notation for the three-phonon transition rates for phonon emission $\left( - \right)$ and absorption $\left( + \right)$ processes as:
\begin{equation}
    \label{eq:pm_transition_rates}
    \mathcal{W}_{\lambda \lambda' \lambda''}^{\pm} 
    = - \frac{\pi \hbar}{4N_{0}} 
    \frac{ \vert \mathbf{\Phi}_{\lambda \left( \pm \lambda' \right) \left(-\lambda''\right)} \vert^{2}}{\omega_{\lambda} \omega_{\lambda'} \omega_{\lambda''}} 
    \delta\left( \omega_{\lambda} \pm \omega_{\lambda'} - \omega_{\lambda''}\right) 
    n_{\lambda}^{0} \left(n_{\lambda'}^{0} + \frac{1}{2} \mp \frac{1}{2} \right) \left( n_{\lambda''}^{0} + 1 \right)
\end{equation}

the linearized three-phonon collision integral $\mathcal{C}^{\text{lin., 3-ph.}} \left( n_{\lambda} \right)$ is simplified as:
\begin{align}
    \label{eq:lin_collision_integral}
    \mathcal{C}^{\text{lin., 3-ph.}} \left( n_{\lambda} \right) 
     = \sum_{\lambda'} \mathcal{T}_{\lambda \lambda'} \tilde{n}_{\lambda'}^{1} 
     = & 
     \sum_{\lambda' \lambda''} 
     \left[
     \mathcal{W}_{\lambda \lambda' \lambda''}^{+} \left( \tilde{n}_{\lambda}^{1} + \tilde{n}_{\lambda'}^{1} - \tilde{n}_{\lambda''}^{1} \right) + \frac{1}{2}
     \mathcal{W}_{\lambda \lambda' \lambda''}^{-} \left( \tilde{n}_{\lambda}^{1} - \tilde{n}_{\lambda'}^{1} - \tilde{n}_{\lambda''}^{1} \right) 
     \right] \\
     = & 
     \mathcal{T}_{\lambda}^{(0)} \tilde{n}_{\lambda}^{1} 
     + \sum_{\lambda' \lambda''} 
     \left[ \mathcal{W}_{\lambda \lambda' \lambda''}^{+} - \frac{1}{2} \mathcal{W}_{\lambda \lambda' \lambda''}^{-} \right]\tilde{n}_{\lambda'}^{1} 
     - \sum_{\lambda' \lambda''} 
     \left[ \mathcal{W}_{\lambda \lambda' \lambda''}^{+} + \frac{1}{2} \mathcal{W}_{\lambda \lambda' \lambda''}^{-} \right]\tilde{n}_{\lambda''}^{1} \nonumber \\
     = & 
     \mathcal{T}_{\lambda}^{(0)} \tilde{n}_{\lambda}^{1} + \sum_{\lambda'} \left[ \sum_{\lambda''} \left( \mathcal{W}_{\lambda \lambda' \lambda''}^{+} - \mathcal{W}_{\lambda \lambda' \lambda''}^{-} - \mathcal{W}_{\lambda' \lambda \lambda''}^{-} \right) \right]\tilde{n}_{\lambda'}^{1} \nonumber \\
    \label{eq:lin_collision_matrix}
     = & 
     \mathcal{T}_{\lambda}^{(0)} \tilde{n}_{\lambda}^{1} + \sum_{\lambda'} \mathcal{T}_{\lambda \lambda'}^{(1)} \tilde{n}_{\lambda'}^{1}
\end{align}
\end{widetext}

Where $\mathcal{T}$ is the three-phonon collision matrix with $\mathcal{T}^{(0)}$ being its diagonal part and $\mathcal{T}^{(1)}$ being its off-diagonal part, given by:
\begin{gather}
    \label{eq:RTA_collision_matrix}
    \mathcal{T}_{\lambda}^{(0)} = \sum_{\lambda' \lambda''} \left[ \mathcal{W}_{\lambda \lambda' \lambda''}^{+} + \frac{1}{2} \mathcal{W}_{\lambda \lambda' \lambda''}^{-} \right] = -\frac{n_{\lambda}^{0} \left( n_{\lambda}^{0} + 1 \right)}{\tau^{\text{3-ph., T}}_{\lambda}} \\
    \label{eq:non_RTA_collision_matrix}
    \mathcal{T}_{\lambda \lambda'}^{(1)} = \sum_{\lambda''} \left( \mathcal{W}_{\lambda \lambda' \lambda''}^{+} - \mathcal{W}_{\lambda \lambda' \lambda''}^{-} - \mathcal{W}_{\lambda' \lambda \lambda''}^{-} \right) = \mathcal{T}_{\lambda' \lambda}^{(1)}
\end{gather}
with $1/\tau^{\text{3-ph., T}}_\lambda = 1/\tau^{\text{3-ph., N}}_\lambda + 1/\tau^{\text{3-ph., U}}_\lambda$ being the total scattering rates for the three-phonon processes, written as a sum of the scattering rates of the N-($1/\tau^{\text{N}}_{\lambda}$) and the U-($1/\tau^{\text{U}}_{\lambda}$) processes. In the above simplification, the following symmetries of the transition rates have been used:
\begin{gather*}
    \mathcal{W}_{\lambda \lambda'' \lambda'}^{-} = \mathcal{W}_{\lambda \lambda' \lambda''}^{-} \\
    \mathcal{W}_{\lambda \lambda'' \lambda'}^{+} = \mathcal{W}_{\lambda' \lambda \lambda''}^{-} \\
    \mathcal{W}_{\lambda' \lambda \lambda''}^{+} = \mathcal{W}_{\lambda \lambda' \lambda''}^{+}
\end{gather*}

\begin{widetext}
\subsubsection{Four-phonon collision matrix}
Similar to three-phonon case, the collision integral for four-phonon scattering processes can be derived as - 
\begin{align}
    \label{eq:4-ph_collosion_integral}
    \mathcal{C}^{\text{4-ph.}} \left( n_{\lambda} \right) 
    = & - \frac{\pi \hbar}{4N_{0}} \sum_{\lambda' \lambda'' \lambda'''} 
    \Bigg[
    \frac{1}{2}
    \frac{ \vert \Phi_{\lambda \lambda' \lambda'' \left( -\lambda'''\right)} \vert^{2} }
    { \omega_{\lambda} \omega_{\lambda'} \omega_{\lambda''} \omega_{\lambda'''} } 
    \delta\left( \omega_{\lambda} + \omega_{\lambda'} + \omega_{\lambda''} - \omega_{\lambda'''} \right) \nonumber \\
    &
    \left( 
    n_{\lambda} n_{\lambda'} n_{\lambda''} \left( n_{\lambda'''} + 1 \right) 
    - \left( n_{\lambda} + 1 \right) \left( n_{\lambda'} + 1 \right) \left( n_{\lambda''} + 1 \right) n_{\lambda'''}
    \right) \nonumber \\
    &
    + \frac{1}{2}
    \frac{ \vert \Phi_{\lambda \lambda' \left( -\lambda''\right) \left( -\lambda'''\right)} \vert^{2} }
    { \omega_{\lambda} \omega_{\lambda'} \omega_{\lambda''} \omega_{\lambda'''} } \delta\left( \omega_{\lambda} + \omega_{\lambda'} - \omega_{\lambda''} - \omega_{\lambda'''} \right) \nonumber \\
    &
    \left( 
    n_{\lambda} n_{\lambda'} \left( n_{\lambda''} + 1 \right) \left( n_{\lambda'''} + 1 \right) 
    - \left( n_{\lambda} + 1 \right) \left( n_{\lambda'} + 1 \right) n_{\lambda''} n_{\lambda'''}
    \right) \nonumber \\
    &
    + \frac{1}{6}
    \frac{ \vert \Phi_{\lambda \left( -\lambda' \right) \left( -\lambda'' \right) \left( -\lambda'''\right)} \vert^{2} }
    { \omega_{\lambda} \omega_{\lambda'} \omega_{\lambda''} \omega_{\lambda'''} } \delta\left( \omega_{\lambda} - \omega_{\lambda'} - \omega_{\lambda''} - \omega_{\lambda'''} \right) \nonumber \\
    &
    \left( 
    n_{\lambda} \left( n_{\lambda'} + 1 \right) \left( n_{\lambda''} + 1 \right) \left( n_{\lambda'''} + 1 \right) 
    - \left( n_{\lambda} + 1 \right) n_{\lambda'} n_{\lambda''} n_{\lambda'''}
    \right)
    \Bigg]
\end{align}
where, the four-phonon scattering matrix element is given by:
\begin{align}
    \label{eq:fourier_transformed_fourth_order_force_constant}
    \mathbf{\Phi}_{\lambda \lambda' \lambda'' \lambda'''} 
    =
    \sum_{l'l''l'''} \sum_{kk'k''k'''} \sum_{\alpha \beta \gamma}  &
    \frac{\Phi_{\alpha \beta \gamma \delta} \left( 0k,l'k',l''k'', l'''k''' \right)}{\sqrt{m_{k} m_{k'} m_{k''} m_{k'''}}} \nonumber \\
    &
    w_{\alpha}\left( \textbf{q}j, k \right) w_{\beta}\left( \textbf{q}'j', k' \right) w_{\gamma}\left( \textbf{q}''j'', k'' \right) w_{\delta}\left( \textbf{q}'''j''', k''' \right) \nonumber \\
    &
    \exp{\left( i \textbf{q}' \cdot \textbf{R}\left( l'\right) \right)}
    \exp{\left( i \textbf{q}'' \cdot \textbf{R}\left( l''\right) \right)}
    \exp{\left( i \textbf{q}''' \cdot \textbf{R}\left( l'''\right) \right)} \nonumber \\
    &
    \Delta\left( \textbf{q}+\textbf{q}'+\textbf{q}''+\textbf{q}''' \right)
\end{align}
with $\Phi_{\alpha \beta \gamma \delta} \left( 0k,l'k',l''k'', l'''k''' \right)$ being the quartic inter-atomic force constants.\\

Linearizing with $n_{\lambda} \approx n_{\lambda}^{0} + n_{\lambda}^{0} \left( n_{\lambda}^{0} + 1 \right) \tilde{n}_{\lambda}^{1}$ and using a short-hand notation $\mathcal{Y}_{\lambda \lambda' \lambda'' \lambda'''}^{\pm\pm}$ for the four-phonon transition rates as above, the linearized four-phonon collision integral becomes: 
\begin{align}
    \mathcal{C}^{\text{lin., 4-ph.}} \left( n_{\lambda} \right) 
     = 
     \sum_{\lambda' \lambda'' \lambda'''} &
     \left[ 
     \frac{1}{2} \mathcal{Y}_{\lambda \lambda' \lambda'' \lambda'''}^{++} \left( \tilde{n}_{\lambda}^{1} + \tilde{n}_{\lambda'}^{1} + \tilde{n}_{\lambda''}^{1} - \tilde{n}_{\lambda'''}^{1} \right) 
     + \frac{1}{2} \mathcal{Y}_{\lambda \lambda' \lambda'' \lambda'''}^{+-} \left( \tilde{n}_{\lambda}^{1} + \tilde{n}_{\lambda'}^{1} - \tilde{n}_{\lambda''}^{1} - \tilde{n}_{\lambda'''}^{1} \right) 
     \right. \nonumber \\
     & \left.
     + \frac{1}{6} \mathcal{Y}_{\lambda \lambda' \lambda'' \lambda'''}^{--} \left( \tilde{n}_{\lambda}^{1} - \tilde{n}_{\lambda'}^{1} - \tilde{n}_{\lambda''}^{1} - \tilde{n}_{\lambda'''}^{1} \right) 
     \right] 
\end{align}
where, the four-phonon transition rates are given by:
\begin{gather*}
    \mathcal{Y}_{\lambda \lambda' \lambda'' \lambda'''}^{++} 
    = 
    - \frac{\pi \hbar}{4N_{0}} 
    \frac{ \vert \mathbf{\Phi}_{\lambda \lambda' \lambda'' \left(-\lambda'''\right)} \vert^{2}}{\omega_{\lambda} \omega_{\lambda'} \omega_{\lambda''} \omega_{\lambda'''}} \delta\left( \omega_{\lambda} + \omega_{\lambda'} + \omega_{\lambda''} - \omega_{\lambda'''}\right) n_{\lambda}^{0} n_{\lambda'}^{0} n_{\lambda''}^{0} \left( n_{\lambda'''}^{0} + 1 \right) \\
    \mathcal{Y}_{\lambda \lambda' \lambda'' \lambda'''}^{+-} 
    = 
    - \frac{\pi \hbar}{4N_{0}} 
    \frac{ \vert \mathbf{\Phi}_{\lambda \lambda' \left( -\lambda'' \right) \left(-\lambda'''\right)} \vert^{2}}{\omega_{\lambda} \omega_{\lambda'} \omega_{\lambda''} \omega_{\lambda'''}} \delta\left( \omega_{\lambda} + \omega_{\lambda'} - \omega_{\lambda''} - \omega_{\lambda'''}\right) n_{\lambda}^{0} n_{\lambda'}^{0} \left(n_{\lambda''}^{0} + 1 \right) \left( n_{\lambda'''}^{0} + 1 \right) \\
    \mathcal{Y}_{\lambda \lambda' \lambda'' \lambda'''}^{--} 
    = 
    - \frac{\pi \hbar}{4N_{0}} 
    \frac{ \vert \mathbf{\Phi}_{\lambda \left( -\lambda' \right) \left( -\lambda'' \right) \left(-\lambda'''\right)} \vert^{2}}{\omega_{\lambda} \omega_{\lambda'} \omega_{\lambda''} \omega_{\lambda'''}} \delta\left( \omega_{\lambda} - \omega_{\lambda'} - \omega_{\lambda''} - \omega_{\lambda'''}\right) n_{\lambda}^{0} \left(n_{\lambda'}^{0} + 1 \right) \left(n_{\lambda''}^{0} + 1 \right) \left( n_{\lambda'''}^{0} + 1 \right)
\end{gather*}

Just as in the three-phonon case, the four-phonon collision matrix $\left( \mathcal{F} \right)$ can be divided into a diagonal $\left( \mathcal{F}^{(0)} \right)$ and an off-diagonal $\left( \mathcal{F}^{(1)} \right)$ part as -
\begin{gather}
    \mathcal{C}^{\text{lin., 4-ph.}} \left( n_{\lambda} \right) 
    = \sum_{\lambda'} \mathcal{F}_{\lambda \lambda'} \tilde{n}_{\lambda'}^{1}
    = 
    \mathcal{F}_{\lambda}^{(0)} \tilde{n}_{\lambda}^{1} + \sum_{\lambda'} \mathcal{F}_{\lambda \lambda'}^{(1)} \tilde{n}_{\lambda'}^{1}
\end{gather}
where,
\begin{gather}
    \mathcal{F}_{\lambda}^{(0)} = \sum_{\lambda' \lambda'' \lambda'''} \left[ \frac{1}{2} \mathcal{Y}_{\lambda \lambda' \lambda'' \lambda'''}^{++} + \frac{1}{2} \mathcal{Y}_{\lambda \lambda' \lambda'' \lambda'''}^{+-} + \frac{1}{6} \mathcal{Y}_{\lambda \lambda' \lambda'' \lambda'''}^{--} \right] = -\frac{n_{\lambda}^{0} \left( n_{\lambda}^{0} + 1 \right)}{\tau_{\lambda}^{\text{4-ph., T}}}
\end{gather}
and,
\begin{align}    
    \mathcal{F}_{\lambda \lambda'}^{(1)} = 
    \sum_{\lambda'' \lambda'''} 
    & 
    \left[ \frac{1}{2} \mathcal{Y}_{\lambda \lambda' \lambda'' \lambda'''}^{++} + \frac{1}{2} \mathcal{Y}_{\lambda \lambda' \lambda'' \lambda'''}^{+-} - \frac{1}{6} \mathcal{Y}_{\lambda \lambda' \lambda'' \lambda'''}^{--} + \frac{1}{2} \mathcal{Y}_{\lambda \lambda'' \lambda' \lambda'''}^{++} - \frac{1}{2} \mathcal{Y}_{\lambda \lambda'' \lambda' \lambda'''}^{+-} \right. \nonumber \\
    & \left. 
    - \frac{1}{6} \mathcal{Y}_{\lambda \lambda'' \lambda' \lambda'''}^{--} - \frac{1}{2} \mathcal{Y}_{\lambda \lambda''' \lambda'' \lambda'}^{++} - \frac{1}{2} \mathcal{Y}_{\lambda \lambda''' \lambda'' \lambda'}^{+-} - \frac{1}{6} \mathcal{Y}_{\lambda \lambda''' \lambda'' \lambda'}^{--} \right]
\end{align}
As discussed in Refs.~\cite{tian_unusual_2018, feng_four-phonon_2017, ravichandran_phonon-phonon_2020}, the diagonal term dominates the four-phonon collision integral for all materials, since the U-processes dominate four-phonon scattering and, unlike the lowest-order three-phonon scattering events, there are no selection rules that apply to these higher-order interactions among four phonons. Hence, we ignore the off-diagonal terms of the four-phonon collision integral in this study.
\end{widetext}

\subsubsection{Solution of the linearized Peierls-Boltzmann equation for phonon transport}
Once the collision matrix is obtained, the linearized Peierls-Boltzmann equation (LPBE), given by:
\begin{align}
    v_{\lambda, x} \frac{\mathrm{d} T}{\mathrm{d} x} 
    \frac{\partial n_{\lambda}^{0}}{\partial T} 
    & \approx \mathcal{C}^{\text{lin.}} \left(n_{\lambda}\right) = \sum_{\lambda'} \mathcal{R}_{\lambda \lambda'} \tilde{n}_{\lambda}^{1}\nonumber\\
    &= \left[\mathcal{T}^{\left(0\right)}_{\lambda} + \mathcal{F}^{\left(0\right)}_{\lambda}\right]\tilde{n}^1_\lambda + \sum_{\lambda'} \mathcal{T}^{\left(1\right)}_{\lambda\lambda'}\tilde{n}^1_{\lambda'}  \\
    &= \mathcal{R}^{\left(0\right)}_\lambda \tilde{n}^1_\lambda + \sum_{\lambda'} \mathcal{R}^{\left(1\right)}_{\lambda\lambda'}\tilde{n}^1_{\lambda'} \label{eq:3+4_LPBE}
\end{align}

\begin{figure}[ht!]
   \centering
    \includegraphics[width=0.95\linewidth]{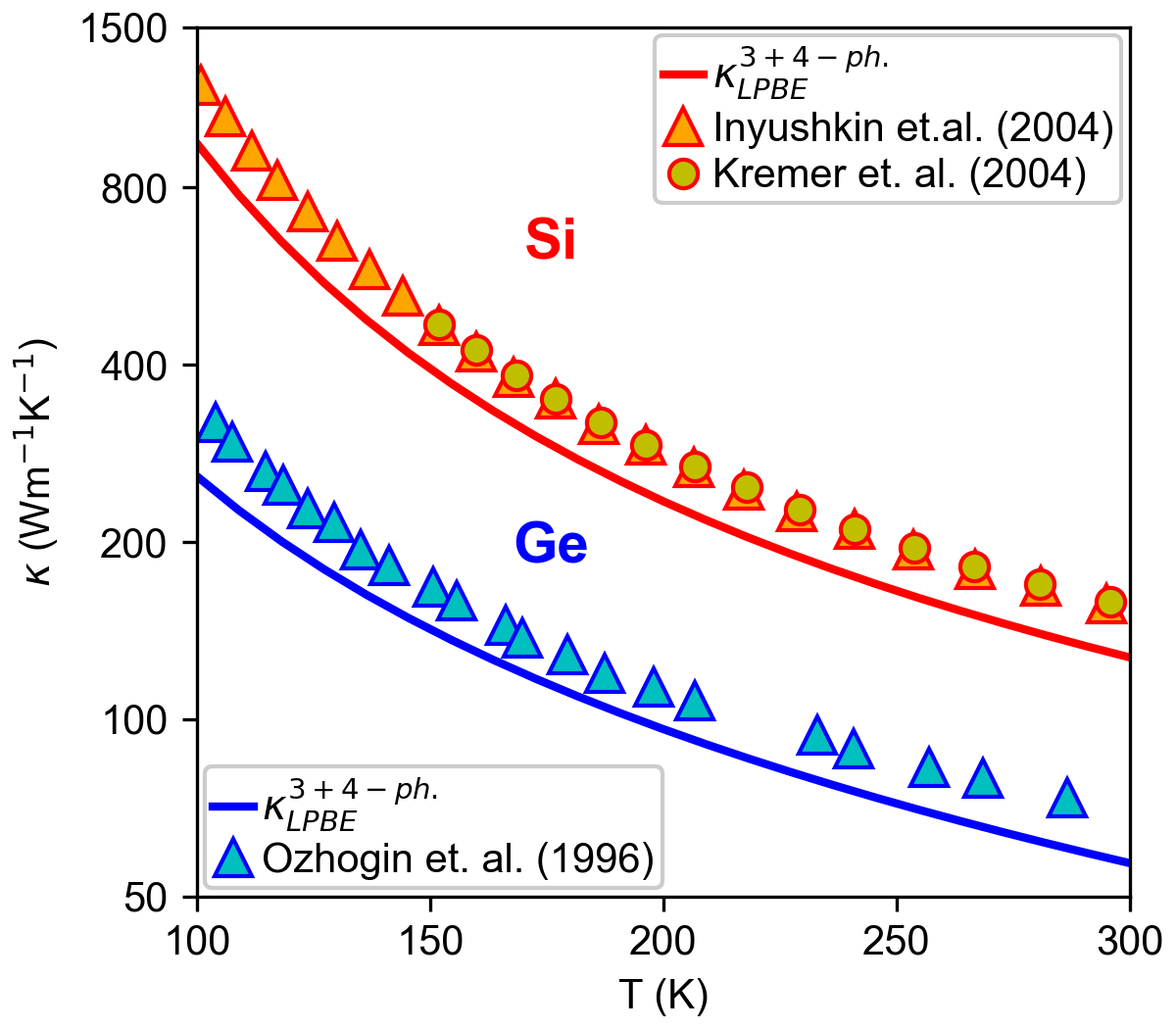}
    \caption{The $\kappa$ of isotopically pure silicon (Si, red) and germanium (Ge, blue), calculated from first-principles with the inclusion of four-phonon scattering. The calculated values are compared with the experimental data from Inyushkin et al.~\cite{inyushkin_isotope_2004} (red triangle), Kremer et al.~\cite{kremer_thermal_2004} (red circle) and Ozhogin et al.~\cite{ozhogin_isotope_1996} (blue triangle). }
    \label{fig:Si_Ge_exp}
\end{figure}
is solved for the unknown variable $\tilde{n}^1_\lambda$. For the calculations in this work, we use a 17$^3$ $\mathbf{q}$-grid and a stabilized bi-conjugate gradient algorithm to solve eq.~\ref{eq:3+4_LPBE}. The harmonic and the anharmonic force constants, the converged parameters for the density functional theory (DFT) calculations to obtain these force constants and the phonon properties such as the phonon frequencies and eigenvectors needed to evaluate eq.~\ref{eq:fourier_transformed_force_constant} are obtained using the methodology described in Ref.~\cite{ravichandran_unified_2018}. These converged parameters of the DFT calculations for all materials except diamond, silicon and germanium are provided in Ref.~\cite{ravichandran_phonon-phonon_2020}. The corresponding converged parameters for silicon, germanium and diamond are provided in tables~\ref{table_DFT_param} and~\ref{Interp_Convergence} below. With these parameters, excellent agreement with available experimental results has been obtained in Ref.~\cite{ravichandran_phonon-phonon_2020} for the $\kappa$ obtained by solving the LPBE from first-principles including three-phonon, four-phonon and phonon-isotope scattering, for silicon carbide and the sixteen different III-V compounds considered in this study. Additionally, it has also been shown in Ref.~\cite{ravichandran_unified_2018} that the calculations for diamond with the same methodology also match well with the experimental data in the literature. For the remaining two materials considered in this study - silicon and germanium, our first-principles prediction of the $\kappa$ agrees well with the experimental data in the literature, as shown in Fig.~\ref{fig:Si_Ge_exp}. Hence, the first-principles $\kappa$ values of all materials considered here to evaluate the effectiveness of the Callaway model are overall in excellent agreement with experimental measurements from different research groups in the literature.

\begin{table*}[htbp]
\caption{First principles computational parameters used in the DFT and the Density Functional Perturbation Theory (DFPT) calculations that produced converged total energy, stress, Born effective charges, and dielectric constants for Si, Ge and diamond. Here, KE is the kinetic energy, wfc. stands for wavefunction, $\rho_{\mathrm{el}}$ is the electron density and ``shifted'' implies a BZ-grid that is not centered at the $\Gamma$-point.}
\begin{center}
{\renewcommand{\arraystretch}{1.5}
\begin{tabular}{c|c|c|c|c}
\hline
Material & KE cutoff (wfc., Ry) & KE cutoff ($\rho_{\mathrm{el}}$, Ry) & $\mathbf{k}$-grid & $\mathbf{q}$-grid (DFPT) \\ \hline
Si & 54 & 216 & 12X12X12-shifted & 8X8X8 \\ \hline
Ge & 35 & 140 & 5X5X5-shifted & 7X7X7 \\ \hline
Diamond & 150 & 600 & 8X8X8 & 8X8X8 \\ \hline
\end{tabular}}
\end{center}
\label{table_DFT_param}
\end{table*}

\begin{table*}[htbp]
\caption{Convergence of $\kappa^{(3)}_{\mathrm{Pure}}$ and $\kappa^{(3+4)}_{\mathrm{Pure}}$ from the LPBE at different temperatures for Si, Ge and Diamond. For $\mathbf{q}$-grids finer than 17$^3$, the three-phonon scattering probabilities ($\mathcal{W}^{\pm}_{\lambda\lambda_1\lambda_2}$) were calculated directly on the finer grid, while the four-phonon scattering is treated under the RTA obtained using a trilinear interpolation scheme from a 17$^3$ $\mathbf{q}$-grid. The $\kappa$'s are in Wm$^{-1}$K$^{-1}$. For each entry, the numbers outside the brackets are $\kappa^{(3+4)}_{\mathrm{Pure}}$ and the numbers within the brackets are $\kappa^{(3)}_{\mathrm{Pure}}$.}
\begin{center}
{\renewcommand{\arraystretch}{1.2}
\begin{tabular}{c|c|c|c|c}
\hline
Material & $\mathbf{q}$-grid & 100 K & 300 K & 750 K \\ \hline
Si  &  17$^{3}$  &  886.35 (932.11)  &  121.38 (134.52)  &  37 (47.8) \\ \hline
	 &  35$^{3}$  &  947.4 (1000.58)  &  128.24 (144.05)  &  39 (51.1) \\ \hline
	 &  51$^{3}$  &  952.11 (1020.66)  &  127.23 (146.21)  &  38.26 (51.8) \\ \hline
  
Ge  &  17$^{3}$  &  244.62 (257.61)  &  55.04 (63.02)  &  16.86 (22.97) \\ \hline
	 &  35$^{3}$  &  256.77 (270.2)  &  57.36 (65.81)  &  17.42 (23.96) \\ \hline
	 &  51$^{3}$  &  258.88 (273.85)  &  56.79 (66.52)  &  17.32 (24.21) \\ \hline
  
Diamond  &  17$^{3}$  &  243364 (244357)  &  3246.69 (3373.85)  &  815.55 (1059.44) \\ \hline
	 &  35$^{3}$  &  184801 (185740)  &  3276.52 (3409.32)  &  824.98 (1071.79) \\ \hline
	 &  51$^{3}$  &  183694 (184674)  &  3286.34 (3420.32)  &  819.52 (1071.65) \\ \hline
\end{tabular}}
\end{center}
\label{Interp_Convergence}
\end{table*}

\subsection{Callaway model calculations}
\subsubsection{Quasimomentum conservation for N-processes}
For the condition on the conservation of phonon quasimomentum ($\hbar\mathbf{q}$) in the presence of N-processes only, Callaway~\cite{callaway_model_1959} imposed the requirement that the rate of change of quasimomentum is zero, i.e.
\begin{align}
    \sum_{j}\int \hbar \textbf{q} \left( \frac{dn_{\lambda}}{dt} \right)_{\text{N}} d\textbf{q} 
    & = 0 \nonumber\\
    \implies \sum_{j}\int \hbar \textbf{q} \mathcal{C}^{\text{N}} \left(n_{\lambda} \right) d\textbf{q} 
    & \approx \sum_{j}\int \hbar \textbf{q} \mathcal{C}^{\text{lin., N}} \left(n_{\lambda} \right) d\textbf{q} \nonumber \\ &
    \approx \sum_{\lambda} \hbar \textbf{q} \mathcal{C}^{\text{lin., N}} \left(n_{\lambda} \right) = 0 \label{eq:n_momentum_conservation}  
\end{align}

Here the superscript N indicates that only N-processes are taken into account. Rewriting eq.~\ref{eq:n_momentum_conservation} in terms of the collision matrix of the N-processes $\left( \mathcal{R}^{\text{N}}_{\lambda\lambda'} \right)$, we get:
\begin{equation}
    \label{eq:momentum_conservation_condition}
    \sum_{\lambda \lambda'} \hbar \textbf{q} \mathcal{R}_{\lambda \lambda'}^{\text{N}} \tilde{n}_{\lambda'}^{1} = 0 
\end{equation}

Furthermore, in the presence of only N-processes, collisions equilibriate $n_{\lambda}$ towards $n^*_\lambda = \frac{1}{\exp \left( \frac{\hbar \omega_{\lambda} }{k_{B}T} + \mathbf{\Lambda}\cdot\textbf{q} \right) - 1}$. Following the usual linearization procedure for $n_{\lambda}^{*}$ about $\mathbf{\Lambda}=\mathbf{0}$, i.e., $n^*_\lambda \approx n_{\lambda}^{0} + n_{\lambda}^{0} \left(n_{\lambda}^{0} + 1 \right)\tilde{n}^{*1}_\lambda $, where $\tilde{n}^{*1}_\lambda = -\textbf{q} \cdot \mathbf{\Lambda}$, we can show, using eq.~\ref{eq:lin_collision_integral}, that $\tilde{n}_{\lambda}^{*1}$ forms a null vector of $\mathcal{R}^{\text{N}}_{\lambda\lambda'}$~\cite{pitaevskii_physical_2012, guyer_solution_1966, krumhansl_thermal_1965, hardy_phonon_1970}, i.e.,
\begin{gather}
    \label{eq:null_vector_N_coll_mat}
    \mathcal{C}^{\text{lin., N}} \left( n_{\lambda}^{*} \right)
    = \sum_{\lambda'} \mathcal{R}_{\lambda \lambda'}^{\text{N}} \tilde{n}_{\lambda'}^{*1} = 0 
\end{gather}

Combining eqs.~\ref{eq:momentum_conservation_condition} and~\ref{eq:null_vector_N_coll_mat}, we get:
\begin{equation}
    \label{eq:modified_momentum_conservation_condition}
    \sum_{\lambda \lambda'} \hbar \textbf{q} \mathcal{R}_{\lambda \lambda'}^{\text{N}} \left( \tilde{n}_{\lambda'}^{1} - \tilde{n}_{\lambda'}^{*1} \right) = 0 
\end{equation}

To get the exact quasimomentum conservation condition used by Callaway~\cite{callaway_model_1959}, the off-diagonal part $\left( \mathcal{R}^{\left( 1 \right), \text{N}}_{\lambda\lambda'} \right)$ is neglected in eq.~\ref{eq:modified_momentum_conservation_condition} to get:
\begin{align*}
    \sum_{\lambda} \hbar \textbf{q} \mathcal{R}_{\lambda}^{\left( 0 \right), \text{N}} \left( \tilde{n}_{\lambda}^{1} - \tilde{n}_{\lambda}^{*1} \right)
    = & \nonumber \\
    - \sum_{\lambda'} \hbar \textbf{q} \frac{n_{\lambda}^{0} \left( n_{\lambda}^{0}+1 \right)}{\tau_{\lambda}^{\text{N}}} & \left( \tilde{n}_{\lambda}^{1} - \tilde{n}_{\lambda}^{*1} \right)
    = 0 
\end{align*}
\begin{gather}
    \label{eq:callaway_condition_for_momentum_conservation}
    \text{i.e.,} - \sum_{\lambda'} \hbar \textbf{q} \left( \frac{ n_{\lambda} - n_{\lambda}^{*} }{\tau_{\lambda}^{\text{N}}} \right) 
    = 0 
\end{gather}

\subsubsection{Thermal conductivity from the Callaway model}
To calculate the $\kappa$ of a material using the Callaway model, we start with eq.~\ref{eq:PBE_callaway}. Introducing the linearized form of $n_{\lambda}^{*}$ ($n^*_\lambda \approx \left[n_{\lambda}^{0} - n_{\lambda}^{0} \left(n_{\lambda}^{0} + 1 \right) \textbf{q} \cdot \mathbf{\Lambda}\right]$) and substituting $\mathbf{\Lambda} = \mathbf{\Theta} \frac{\mathrm{d} T}{\mathrm{d} x}$, where $\mathbf{\Theta}$ is the phonon mobility, the LPBE becomes:
\begin{widetext}
\begin{align}
    v_{\lambda, x} \frac{\partial n_{\lambda}^{0}}{\partial T} \frac{\mathrm{d} T}{\mathrm{d} x} 
    = & 
    - \frac{n_{\lambda} - n_{\lambda}^{0}}{\tau_{\lambda}^{\text{T}}}
    - \frac{n_{\lambda}^{0} \left( n_{\lambda}^{0} + 1 \right)}{\tau_{\lambda}^{\text{N}}}
    \textbf{q} \cdot \mathbf{\Theta} \frac{\mathrm{d} T}{\mathrm{d} x} \nonumber \\
    \text{i.e., } n_{\lambda} - n_{\lambda}^{0}
    = & 
    - v_{\lambda, x} \tau_{\lambda}^{\text{T}} \frac{\partial n_{\lambda}^{0}}{\partial T} \frac{\mathrm{d} T}{\mathrm{d} x} - \frac{\tau_{\lambda}^{\text{T}}}{\tau_{\lambda}^{\text{N}}} n_{\lambda}^{0} \left( n_{\lambda}^{0} + 1 \right) \textbf{q} \cdot \mathbf{\Theta} \frac{\mathrm{d} T}{\mathrm{d} x} 
\end{align}

Using the linearized form of $n^*_\lambda$ in the quasimomentum conservation condition for the N-processes (eq.~\ref{eq:callaway_condition_for_momentum_conservation}), we get:
\begin{gather}
    \sum_{\lambda} \hbar \textbf{q} \left( \frac{n_{\lambda} - n^{*}_{\lambda}}{\tau^{\text{N}}_{\lambda}} \right) 
    \approx \sum_{\lambda} \hbar \textbf{q} \left( \frac{n_{\lambda} - n_{\lambda}^{0} }{\tau_{\lambda}^{\text{N}}} + \frac{n_{\lambda}^{0} \left(n_{\lambda}^{0}+1 \right) \textbf{q} \cdot \mathbf{\Lambda}}{\tau_{\lambda}^{\text{N}}} \right) = 0 \nonumber \\
    \implies
    \sum_{\lambda} \hbar \textbf{q} \left( - \frac{\tau_{\lambda}^{\text{T}}}{\tau_{\lambda}^{\text{N}}} + 1 \right) \frac{n_{\lambda}^{0} \left(n_{\lambda}^{0}+1 \right) \textbf{q} \cdot \mathbf{\Lambda}}{\tau_{\lambda}^{\text{N}}}
    = \sum_{\lambda} \hbar \textbf{q} \left( v_{\lambda, x} \frac{\tau_{\lambda}^{\text{T}}}{\tau_{\lambda}^{\text{N}}} \frac{\partial n_{\lambda}^{0}}{\partial T} \frac{\mathrm{d} T}{\mathrm{d} x} \right)
    \nonumber \\
    \implies
    \sum_{\lambda} \textbf{q} n_{\lambda}^{0} \left(n_{\lambda}^{0}+1 \right) \textbf{q} \cdot \mathbf{\Lambda} \frac{\tau_{\lambda}^{\text{T}}}{\tau_{\lambda}^{\text{U}} \tau_{\lambda}^{\text{N}}} 
    = \sum_{\lambda} \textbf{q} \left( v_{\lambda, x} \frac{\tau_{\lambda}^{\text{T}}}{\tau_{\lambda}^{\text{N}}} \frac{\partial n_{\lambda}^{0}}{\partial T} \right) \frac{\mathrm{d} T}{\mathrm{d} x}
    \nonumber \\
    \implies
    \mathbf{\Lambda} 
    = \sum_{\lambda} \hat{\textbf{q}} \left( v_{\lambda, x} \frac{\tau_{\lambda}^{\text{T}}}{\tau_{\lambda}^{\text{N}}} \frac{\partial n_{\lambda}^{0}}{\partial T} \right) \frac{\mathrm{d} T}{\mathrm{d} x} = \mathbf{\Theta} \frac{\mathrm{d} T}{\mathrm{d} x}
\end{gather}
\end{widetext}
Here, $\hat{\textbf{q}} = \Pi^{-1}\textbf{q}$, where $\Pi$ given by:
\begin{gather*}
    \Pi = \sum_{\lambda} \textbf{q} \otimes \textbf{q} n_{\lambda}^{0}(n_{\lambda}^{0}+1) 
    \frac{\tau_{\lambda}^{\text{T}}}{\tau_{\lambda}^{\text{U}} \tau_{\lambda}^{\text{N}}}
\end{gather*}

Next, the heat flux $J_{x}$ and hence the thermal conductivity $\kappa$ are calculated as:
\begin{align}
    J_{x} = & \frac{1}{\Omega} \sum_{\lambda} \hbar \omega_{\lambda} v_{\lambda, x} \left( n_{\lambda} - n_{\lambda}^{0} \right) \nonumber \\
    - \kappa \frac{\partial T}{\partial x} = & - \frac{1}{\Omega} \sum_{\lambda} \hbar \omega_{\lambda} v_{\lambda, x}^{2} \tau_{\lambda}^{\text{T}} \frac{\partial n_{\lambda}^{0}}{\partial T} \frac{\mathrm{d} T}{\mathrm{d} x} \nonumber \\ & - \frac{1}{\Omega} \sum_{\lambda} \hbar \omega_{\lambda} \frac{\tau_{\lambda}^{\text{T}}}{\tau_{\lambda}^{\text{N}}} n_{\lambda}^{0} \left( n_{\lambda}^{0} + 1 \right) \textbf{q} \cdot \mathbf{\Lambda} \nonumber \\
    \kappa = & \frac{1}{\Omega} \sum_{\lambda} \hbar \omega_{\lambda} v_{\lambda, x}^{2} \tau_{\lambda}^{\text{T}} \frac{\partial n_{\lambda}^{0}}{\partial T} \nonumber \\ & + \frac{1}{\Omega} \sum_{\lambda} \hbar \omega_{\lambda} v_{\lambda, x} \frac{\tau_{\lambda}^{\text{T}}}{\tau_{\lambda}^{\text{N}}} n_{\lambda}^{0} \left( n_{\lambda}^{0} + 1 \right) \textbf{q} \cdot \mathbf{\Theta}
\end{align}

\section*{Appendix - II: Performance of the Callaway model for aluminum antimonide (AlSb)}
\begin{figure*}[!ht]
\includegraphics[width = 0.99\linewidth]{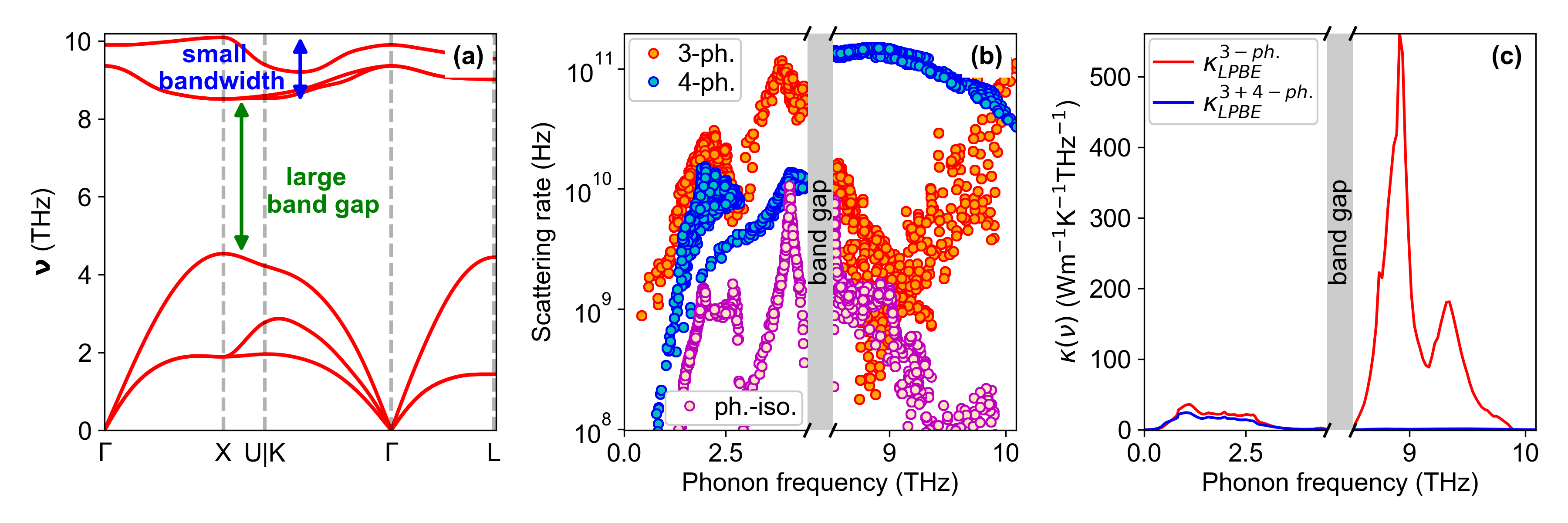}
\centering
\caption{\label{fig:AlSb_disp_scat_rate_plot} (a) Phonon dispersions for aluminum antimonide (AlSb), showing the key features that activate the AAO and AOO selection rules on phonon scattering processes. (b) Scattering rates vs. phonon frequency for AlSb at 300 K (c) Spectral contribution to $\kappa$ with (blue) and without (red) the inclusion of four-phonon scattering in isotopically pure AlSb at 300 K.}
\end{figure*}

In AlSb, the three-phonon scattering rates are low for the optic phonons due to the simultaneous activation of AOO and AAO selection rules (A: acoustic, O: optic) as discussed in Ref.~\cite{ravichandran_phonon-phonon_2020}. These low scattering rates occur in the region of the Brillouin zone, where the group velocities are non-zero, as can be seen in Fig.~\ref{fig:AlSb_disp_scat_rate_plot}~(a) and~(b) and discussed in Ref.~\cite{ravichandran_phonon-phonon_2020}, thus resulting in a large contribution of optic phonons to total $\kappa$, when only three-phonon scattering processes are considered. On the other hand, for the acoustic phonons, the AAO selection rule is alone activated; hence, their three-phonon scattering rates are not as low as for the optic phonons, and so, their contribution to $\kappa$ is relatively small [Fig.~\ref{fig:AlSb_disp_scat_rate_plot} (c)]. The features in the phonon dispersions that activate these selection rules are highlighted in Fig.~\ref{fig:AlSb_disp_scat_rate_plot} (a).

However, since the four-phonon scattering rates for the optic phonons in AlSb are much larger than their lower-order three-phonon counterparts, as shown in Fig.~\ref{fig:AlSb_disp_scat_rate_plot}~(b), the acoustic phonons become the primary heat carriers when four-phonon scattering is included in the calculations, and the total $\kappa$ including three- and four-phonon scattering will be much smaller than that including only three-phonon scattering [Fig.~\ref{fig:AlSb_disp_scat_rate_plot} (c)].

\begin{figure*}[!ht]
\includegraphics[width = 0.99\linewidth]{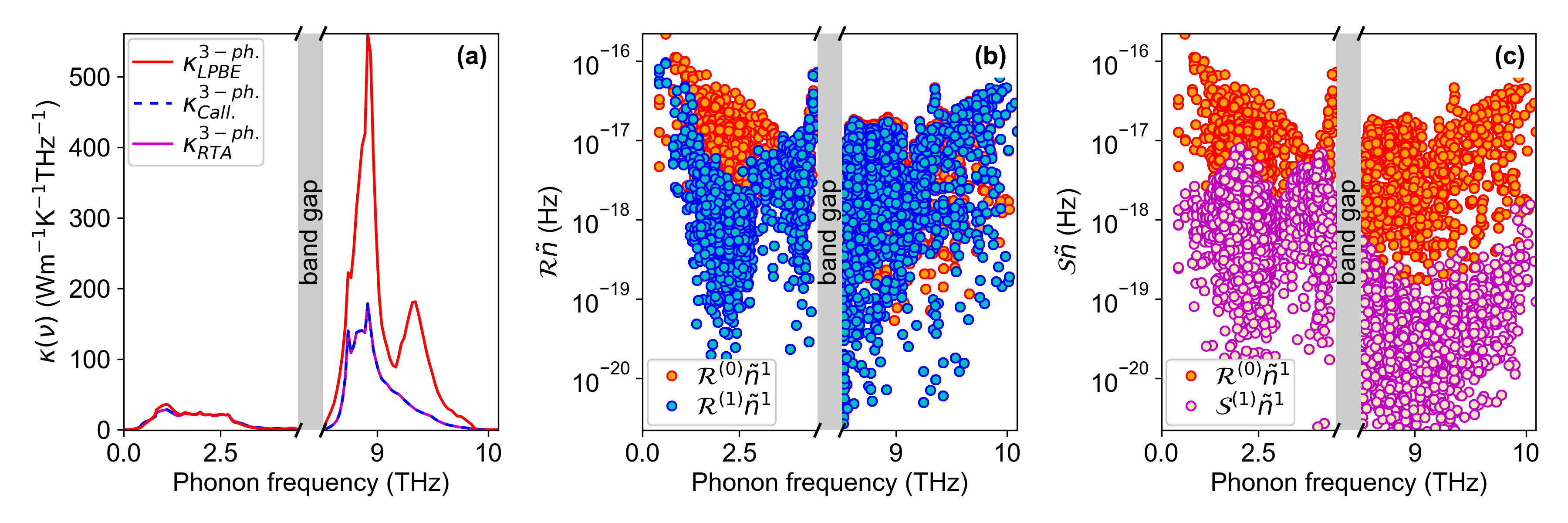}
\centering
\caption{\label{fig:AlSb_rta_validation_plot} (a) Spectral contribution to $\kappa$ without the inclusion of four-phonon scattering in isotopically pure AlSb at 300 K from the solutions of the RTA (pink solid line), the complete LPBE (red solid line) and the Callaway model (blue dashed line). In Figs. (b) and (c), we compare the diagonal and the off-diagonal terms of collision integrals $\mathcal{C}^{\text{lin.}} \left( \tilde{n}^{1} \right)$ and  $\mathcal{C}^{\text{Call.}} \left( \tilde{n}^{1} \right)$ respectively.}
\end{figure*} 
To check the suitability of the Callaway model for AlSb, the two conditions listed in the main text must be considered. As shown in Fig.~\ref{fig:AlSb_rta_validation_plot} (a), when four-phonon scattering is ignored, the RTA significantly under-predicts the complete solution of the LPBE only for the optic phonons in AlSb. This observation is also reflected in Fig.~\ref{fig:AlSb_rta_validation_plot} (b), where the diagonal terms $\mathcal{R}^{\left(0\right)}_\lambda \tilde{n}^1_\lambda$ of $\mathcal{C}^{\text{lin.}}\left(n_\lambda\right)$ are much larger than the off-diagonal terms $\sum_{\lambda\lambda'}\mathcal{R}^{\left(1\right)}_{\lambda\lambda'}\tilde{n}^1_{\lambda'}$ for the acoustic phonons, but are comparable to each other for the optic phonons. On the other hand, the solution from the Callaway model shows much smaller off-diagonal terms in the Callaway collision integral for all phonons, and particularly for the optic phonons. Due to this large difference in the off-diagonal terms of the collision integrals in the LPBE and the Callaway model, the latter fails dramatically to capture the complete solution of the LPBE for AlSb when only three-phonon scattering is included. When four-phonon scattering is also included, the acoustic phonons become the primary heat carriers in AlSb as discussed before. Since the diagonal terms of the collision integral from LPBE for the acoustic phonons are much larger than their off-diagonal counterparts [Fig.~\ref{fig:AlSb_rta_validation_plot} (b)], the predicted $\kappa$ from the RTA, the LPBE and the Callaway approximation overlap for AlSb, when four-phonon scattering is included in the calculations.

\nocite{*}
\newpage
\bibliography{references}

\end{document}